\documentclass[]{aastex631}
\shorttitle{AASTeX v6.3.1 Sample article} 
\shortauthors{Inoue et al.2023}
\graphicspath{{./}{figures/}}

\usepackage{CJK}
\usepackage{bm}
\usepackage{multirow}

\begin{document}
\begin{CJK*}{UTF8}{ipxm}

\title{Modeling of Thermal Emission from ULX Pulsar Swift J0243.6+6124 
with General Relativistic Radiation MHD simulations}

\author[0000-0002-0700-2223]{Akihiro Inoue}
\affiliation{Center for Computational Sciences, University of Tsukuba, 1-1-1 Ten-nodai, Tsukuba, Ibaraki 305-8577, Japan}

\author[0000-0002-2309-3639]{Ken Ohsuga}
\affiliation{Center for Computational Sciences, University of Tsukuba, 1-1-1 Ten-nodai, Tsukuba, Ibaraki 305-8577, Japan}
\email{akihiro@ccs.tsukuba.ac.jp}


\author[0000-0003-0114-5378]{Hiroyuki R. Takahashi}
\affiliation{Department of Natural Sciences, Faculty of Arts and Sciences, Komazawa University, Tokyo 154-8525, Japan}

\author[0000-0003-3640-1749]{Yuta Asahina}
\affiliation{Center for Computational Sciences, University of Tsukuba, 1-1-1 Ten-nodai, Tsukuba, Ibaraki 305-8577, Japan}



\begin{abstract}
We perform general relativistic 
radiation magnetohydrodynamics (MHD) simulations 
of super-Eddington accretion flows 
around a neutron star with a dipole magnetic field
for modeling the galactic ultra-luminous X-ray source (ULX) 
exhibiting X-ray pulsations, Swift J0243.6+6124. 
Our simulations show the accretion columns near the magnetic poles, 
the accretion disk outside the magnetosphere, 
and the outflows from the disk. 
It is revealed that the effectively optically thick outflows, 
consistent with the observed thermal emission 
at $\sim10^7$ K, 
are generated 
if the mass accretion rate is much higher 
than the Eddington rate $\dot{M}_{\rm Edd}$ 
and the magnetospheric radius is smaller
than the spherization radius.
In order to explain the blackbody radius
($\sim 100-500$ km)
without contradicting the reported spin period ($9.8~{\rm s}$)
and spin-up rate ($\dot{P}=-2.22\times10^{-8}~{\rm s~s^{-1}}$),
the mass accretion rate of $(200-1200)\dot{M}_{\rm Edd}$ is required.
Since the thermal emission was detected in two observations 
with $\dot{P}$ of
$-2.22\times10^{-8}~{\rm s~s^{-1}}$ and
$-1.75\times10^{-8}~{\rm s~s^{-1}}$
but not in another with 
$\dot{P}=-6.8 \times10^{-9}~{\rm s~s^{-1}}$,
the surface magnetic field strength of the neutron star in 
Swift J0243.6+6124 is estimated to be
between $3\times10^{11}~{\rm G}$ 
and $4\times10^{12}~{\rm G}$.
From this restricted range of magnetic field strength,
the accretion rate would be 
$(200-500)\dot{M}_{\rm Edd}$
when the thermal emission appears 
and 
$(60-100)\dot{M}_{\rm Edd}$ 
when it is not detected.
Our results support the hypothesis 
that the super-Eddington phase 
in the 2017-2018 giant outburst of Swift J0243.6+6124
is powered by highly super-Eddington accretion flows 
onto a magnetized neutron star.
\end{abstract}



{\section{Introduction} \label{sec:intro}}

Ultra-Luminous X-ray sources (ULXs) are bright X-ray compact objects, 
whose X-ray luminosities exceed the Eddington luminosity 
for stellar mass black holes $\sim10^{39}~{\rm erg~s^{-1}}$, 
and have been discovered in off-nuclear regions of our galaxy and nearby galaxies
\citep[see reviews by][]{Kaaret2017,Fabrika2021}.
Recently, several ULXs have been shown to 
have coherent pulsations at periods, $\sim1-10~\rm s$ 
\citep[e.g.,][]{Bachetti2014,Furst2016,Israel2017a,Israel2017b}.
These objects are called ULX Pulsars (ULXPs).
It is widely believed that 
these pulsations can be observed
if the radiation from the vicinity of the neutron star (NS)
periodically changes via rotation of the NS.
In this case, the ULXP is powered by a super-Eddington 
accretion onto the magnetized NS 
since the luminosity of the ULXP exceeds the Eddington luminosity
for the NS, $L_{\rm Edd}$.

The super-Eddington accretion flows around the magnetized NS
consist of the accretion disk, 
the outflows, and the accretion columns.
The super-Eddington accretion disk is 
truncated at a certain radius,
called the truncation radius,
by the magnetic fields of the NS, 
provided that the magnetic field strength of the NS is 
high enough to prevent disk accretion.
It is considered that radiatively driven outflows
are launched from the super-Eddington accretion disk,
which forms on the outside of the truncation radius,
since the vertical radiation pressure force is larger than
the vertical gravitational force \citep{Shakura1973,Poutanen2007}.
The gas moves along the magnetic field lines inside 
the truncation radius.
Then, column accretion flows, called the accretion columns,
are formed near the magnetic poles of the NS.
The radiative luminosity at the base of the accretion columns
exceeds the Eddington luminosity
\citep{Basko1976,Mushtukov2015,Kawashima2016}.
If the magnetic axis is misaligned with the rotation axis of the NS,
then the pulsation exceeding the Eddington luminosity can be observed
via the precession of the accretion columns
\citep{Mushtukov2018,Inoue2020}.
This inflow-outflow structure around the magnetized NS 
has been shown by general relativistic radiation magnetohydrodynamics (GR-RMHD) 
simulations of super-Eddington accretion flows 
onto a NS with a dipole magnetic field
\citep[][]{Takahashi2017,Abarca2021}.

Recent observations suggested that outflows exist in ULXs 
\citep[see e.g.,][]{Pinto2016} and ULXPs \citep[see e.g.,][]{Kosec2018}. 
In the galactic ULXP, Swift J0243.6+6124, the outflows, 
as well as a relativistic jet \citep{Eijnden2018}, 
were reported \citep[see e.g.,][]{Tao2019,Eijnden2019}.
\citet{Tao2019} showed that the radiation spectrum of Swift J0243.6+6124,
at the peak of the 2017-2018 giant outburst,
can be explained by the thermal blackbody 
at $0.4-0.6 ~{\rm keV}\sim10^7~{\rm K}$, 
with a radius of $100-500 ~\rm km$ \citep[however, see also][]{Jaisawal2019}.
This thermal emission is considered to originate from 
the optically thick outflows.
It is not known whether 
the outflows appearing in the numerical simulations 
mentioned above can explain such thermal emission. 
Although a semi-analytical model was proposed 
to account for the effects of the outflows \citep{Mushtukov2019,Chashkina2019}, 
it remains to be unclear whether such powerful outflows
can occur or not.

In this paper, 
we perform GR-RMHD simulations of 
super-Eddington accretion flows onto a NS 
with a dipole magnetic field for modeling of Swift J0243.6+6124. 
The dependence of the magnetospheric and outflow structure on 
the surface magnetic field strength at the NS surface $B_{\rm NS}$ 
and the mass accretion rate $\dot{M}_{\rm in}$,
which is not investigated in previous studies
\citep{Takahashi2017,Abarca2021},
is reported.
It is revealed that
the super-Eddington accretion onto 
the magnetized NS 
reproduces the thermal emission 
observed in this object.
We also restrict $B_{\rm NS}$ and $\dot{M}_{\rm in}$ based on
the condition of the thermal emission.
The paper is organized as follows:
we will describe the numerical methods  
in Section \ref{sec:methods} and then
present the results in Section \ref{sec:results}. 
Section \ref{sec:discussion} is devoted to 
the discussion.
Finally, we give our conclusion in the final section.

{\section{Numerical Methods} \label{sec:methods}}

In this study,
we numerically solve the GR-RMHD equations 
in Schwarzschild polar coordinates $(t,r,\theta,\phi)$,
assuming axisymmetry with respect to $\theta=0,\pi$.
We use the GR-RMHD simulation code, {\tt\string UWABAMI}
\citep{Takahashi2017}.
The moment formalism is adopted to describe the radiation field
\citep{Thorne1981}.
M1 closure is used as the closure relation
\citep{Levermore1984,Kanno2013,Sadowski2013}.
Using the closure, 
the radiation fields can be updated by solving the 0-th and 1st 
moment equations of the radiation 
without solving the radiative transfer equation.
Hereafter, Greek suffixes and Latin suffixes 
represent spacetime components and space components, respectively.
The speed of light $c$ and the gravitational constant $G$ 
are set to $1$ unless otherwise noted.

\subsection{Basic equations}
The basic equations are as follows
\citep[see e.g.,][]{Takahashi2018},
the mass conservation law,
\begin{eqnarray}
    \left(\rho u^\mu\right)_{;\mu}=0,
    \label{eq:mass_cons}
\end{eqnarray}
the energy-momentum conservation laws for magnetohydrodynamics (MHD),
\begin{eqnarray}
    \left(T^{\mu\nu}\right)_{;\nu}=G^\mu,
    \label{eq:momentum_cons_gas}
\end{eqnarray}
the energy-momentum conservation laws for the radiation field,
\begin{eqnarray}
    \left(R^{\mu\nu}\right)_{;\nu}=-G^\mu,
    \label{eq:momentum_cons_rad}
\end{eqnarray}
the induction equations,
\begin{eqnarray}
    \partial_t\left(\sqrt{-g}B^i\right)+
    \partial_j\left\{\sqrt{-g}\left(b^i u^j-b^j u^i\right)\right\}=0,
    \label{eq:induction_eq}
\end{eqnarray}
where
$\rho$ is the proper mass density, 
$u^\mu$ is the four-velocity of the gas,
$B^i$ is the magnetic field vector in 
the laboratory frame, 
$b^\mu$ is the magnetic four-vector in the fluid frame,
and $g$ is the determinant of the metric, $g=\det{(g_{\mu\nu})}$.
The energy momentum tensor of ideal MHD $T^{\mu\nu}$
consists of that of fluid ${T_{\rm gas}}^{\mu\nu}$ 
and electromagnetic field $M^{\mu\nu}$:
\begin{eqnarray}
    {T_{\rm gas}}^{\mu\nu}
    &=&(\rho+e+p_{\rm gas})u^{\mu}u^{\nu}
     +p_{\rm gas} g^{\mu\nu},
     \label{eq:Tmunu_gas}\\
    {M}^{\mu\nu}
    &=&b^2u^{\mu}u^{\nu}+p_{\rm mag} g^{\mu\nu}-b^\mu b^\nu,
    \label{eq:_Mmunu}
\end{eqnarray}
where
$e$ is the internal energy density,
$p_{\rm gas}=(\Gamma-1)e$ is the gas pressure ($\Gamma=5/3$),
and $p_{\rm mag}=b^2/2$ is the magnetic pressure in the fluid frame.
The energy momentum tensor of the radiation field in the M1 formalism is given by
\citet{Sadowski2013},
\begin{eqnarray}
    R^{\mu\nu}
    =\left(\bar{E}+p_{\rm rad}\right){u_{\rm R}}^{\mu}{u_{\rm R}}^{\nu}
    +\frac{1}{3}\bar{E}g^{\mu\nu}.
    \label{eq:Rmunu}
\end{eqnarray}
Here, $\bar{E}$, $p_{\rm rad}=\bar{E}/3$, and ${u_{\rm R}}^{\mu}$ are 
the radiation energy density, 
the radiation pressure in the radiation rest frame,
and the four-velocity of the radiation rest frame, respectively.
The radiation four-force $G^\mu$, 
which describes the interaction between the ideal MHD and radiation field,
is defined as
\begin{eqnarray}
    {G}^\mu =
    &-&{\rho} {\kappa}_{\rm abs}
    \left(R^{\mu\alpha} u_\alpha+4\pi \hat{B} u^\mu\right)\nonumber\\
    &-&{\rho}{\kappa}_{\rm sca}
    \left(R^{\mu\alpha} u_\alpha
    +R^{\alpha\beta} u_\alpha u_\beta u^\mu\right)
    +{G_{\rm comp}}^\mu,
\end{eqnarray}
where $\kappa_{\rm abs}$ and $\kappa_{\rm sca}$
are the opacity for free-free absorption 
and isotropic electron scattering, respectively:
\begin{eqnarray}
    \kappa_{\rm abs}&=&6.4\times 10^{22}\rho T_{\rm e}^{-3.5}~~~[{\rm cm^2~g^{-1}}],
    \label{eq:ffabs}\\
    \kappa_{\rm sca}&=&0.4~~~[{\rm cm^2~g^{-1}}].
    \label{eq:electron_sca}
\end{eqnarray}
Here, $T_{\rm e}$ is the electron temperature.
The blackbody intensity is given by 
$\hat{B}= a{{T}_{\rm e}}^4/4\pi$,
where $a$ is the radiation constant.
In this study, 
we consider the thermal comptonization defined as follows
\citep{Fragile2018,Utsumi2022}：
\begin{eqnarray}
    {G_{\rm comp}}^\mu
    =-\kappa_{\rm sca}\rho\hat{E}
    \frac{4k(T_{\rm e}-T_{\rm r})}{m_{\rm e}}u^\mu.
    \label{eq:thermal_comp}
\end{eqnarray}
Here, 
$\hat{E}$ is the radiation energy density in the fluid frame, 
$T_r=(\hat{E}/a)^{1/4}$ is the radiation temperature and 
$m_{\rm e}$ is the electron rest mass.
We take $T_{\rm e}=T_{\rm g}$ for simplicity,
where $T_{\rm g}$ is the gas temperature.
The gas temperature can be derived from
\begin{eqnarray}
    T_{\rm g}=\frac{\mu_{\rm w} m_{\rm p} p_{\rm gas}}{\rho k},
\end{eqnarray}
where $m_{\rm p}$ is the proton mass,
$k$ is the Boltzmann constant,
and $\mu_{\rm w}=0.5$ is the mean molecular weight.
We consider the mean-field
dynamo in our simulations
\citep{Sadowski2015}.

{\subsection{Numerical models}\label{sec:numrical_model}}

\begin{deluxetable*}{p{30mm}ccccc}[htb]
\tablenum{1}
\tablecaption{Parameters for different models\label{tab:table1}}
\tablewidth{0pt}
\tablehead{
\colhead{Parameter}
&\colhead{${B_{\rm NS}}$}
&\colhead{${\rho_0}$}
&\colhead{$t_{\rm end}$}
&\colhead{$t_{\rm eq}$} 
&\colhead{$(N_r,N_\theta)$}\\
\colhead{Unit}
&\colhead{$[\rm G]$}
&\colhead{$[\rm g~cm^{-3}]$}
&\colhead{$[t_{\rm g}]$}
&\colhead{$[t_{\rm g}]$}
&\colhead{}
}
\startdata
{\tt\string B3e9D001}    & ~~~$3.3\times10^{9}$~~~  & 0.01  & $50000$ & $30000$  & 
(592, 412)\\
{\tt\string B3e9D01\_np} & ~~~$3.3\times10^{9}$~~~  & 0.1   & $50000$ & $40000$  & 
(592, 412)\\
{\tt\string B1e10D001}    & ~~~$10^{10}$~~~         & 0.01  & $50000$ & $30000$  & 
(592, 512)\\
{\tt\string B1e10D002}    & ~~~$10^{10}$~~~         & 0.021 & $50000$ & $30000$  & 
(592, 512)\\
{\tt\string B1e10D004}    & ~~~$10^{10}$~~~         & 0.046 & $50000$ & $35000$  & 
(592, 512)\\
{\tt\string B1e10D01}     & ~~~$10^{10}$~~~         & 0.1   & $50000$ & $35000$  & 
(592, 412)\\
{\tt\string B1e10D1\_np}  &  ~~~$10^{10}$~~~        & 1     & $50000$ & $40000$  & 
(592, 512)
\\
{\tt\string B3e10D001}    & ~~~$3.3\times10^{10}$~~~& 0.01  & $50000$ & $40000$  & 
(732, 512)\\
{\tt\string B3e10D01}     & ~~~$3.3\times10^{10}$~~~& 0.1   & $50000$ & $35000$  & 
(732, 512)\\
{\tt\string B3e10D02}     & ~~~$3.3\times10^{10}$~~~& 0.21  & $50000$ & $40000$  & 
(592, 512)\\
{\tt\string B3e10D04}     & ~~~$3.3\times10^{10}$~~~& 0.46  & $50000$ & $40000$  & 
(592, 512)\\
{\tt\string B3e10D1}      & ~~~$3.3\times10^{10}$~~~&  1    & $50000$ & $40000$  & 
(592, 412)\\
{\tt\string B1e10D001\_a} &  ~~~$10^{10}$~~~        & 0.01  & $40000$ & $25000$   & 
(732, 512)\\
{\tt\string B1e10D001\_b} &  ~~~$10^{10}$~~~        & 0.01  & $40000$ & $30000$   & 
(592, 412)\\
{\tt\string B1e10D001\_c} &  ~~~$10^{10}$~~~        & 0.01  & $40000$ & $30000$   & 
(592, 326)
\enddata
\tablecomments{
$B_{\rm NS}$, $\rho_0$, and $(N_r,N_\theta)$ are
the surface magnetic field strength of the NS,
the maximum gas density of the initial torus,
and numerical grid points, respectively.
The simulation continues until $t=t_{\rm end}$.
The mass outflow rate is almost constant after $t=t_{\rm eq}$.
The model without accretion columns 
is denoted by {\tt\string _np} at the end of the model name.
}
\end{deluxetable*}

In this study,
we take $M_{\rm NS}=1.4M_\odot$ and $r_{\rm NS}=10~\rm km$,
where $M_{\rm NS}$ and $r_{\rm NS}$ 
are the mass and radius of the NS,
and consider the NSs with dipole magnetic fields \citep{Wasserman1983}.
The dipole magnetic field strength at the NS surface $B_{\rm NS}$
applied in this study, 
which is relatively weaker than the typical value observed in the X-ray pulsar,
$10^{11-13}~{\rm G}$ (see Section \ref{sec:future}),
is described in Table \ref{tab:table1}.
The table also reports 
the maximum gas density of the initial torus $\rho_0$,
the end time of the simulation $t_{\rm end}$,
the time after which the mass outflow rate is almost constant $t_{\rm eq}$,
and numerical grid points $(N_r, N_\theta)$.
The light-crossing time 
for the gravitational radius of the NS, 
$r_{\rm g}=M_{\rm NS}=2.1~{\rm km}$,
is denoted by $t_{\rm g}$.
The computational domain consists of 
$r=[r_{\rm NS},r_{\rm out}]$,
where $r_{\rm out}=2100~{\rm km}$, 
and $\theta=[0,\pi]$.
The size of the radial grid 
exponentially increases with $r$,
and the size of the polar grid is uniform \citep{Takahashi2017}.
Multipole components of the NS magnetic field are not considered.
GR-RMHD simulations considering the multipole components 
are left as an important future work,
although the GR-MHD simulations employing quadrupole fields 
were recently performed by \citet{Das2022}.
The rotation of the NSs is ignored since
the rotation period of the NSs observed in ULXPs, $1-10$ s, 
is much longer than the rotational timescale of the accretion disk, 
$\sim10^{-2}~{\rm s}$,
even within $r\sim100~{\rm km}$.
There remains the possibility that ULX is a millisecond pulsar \citep{Kluzniak2015}, 
which we will discuss later.
We assume that 
the magnetic axis coincides with 
the rotation axis of the accretion disk.
The convergence of the simulation results is 
confirmed using models 
{\tt\string B1e10D001}, {\tt\string B1e10D001\_a}, 
{\tt\string B1e10D001\_b}, and {\tt\string B1e10D001\_c}.
These models have the same initial parameters except for the resolution.

\begin{figure}[tb]
\centering
\includegraphics[width=85mm]{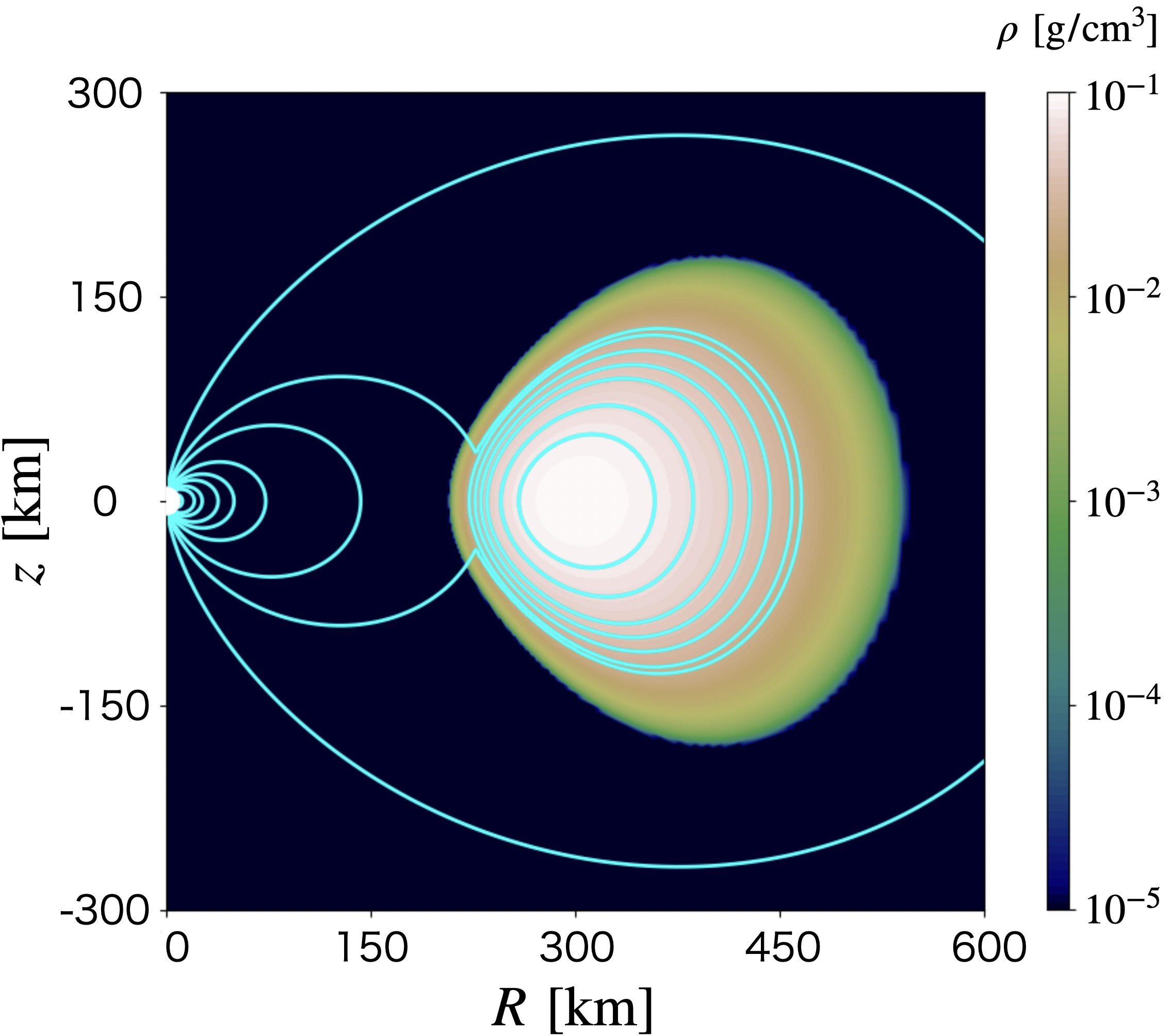}
\caption{
Snapshot of density for 
initial equilibrium torus in the case of model {\tt\string B1e10D01}.
Solid lines represent magnetic field lines.
\label{fig:figure1}}
\end{figure}

We initially set the equilibrium torus given by \citet{Fishbone1976},
as can be seen in Figure \ref{fig:figure1},
where $(R,z)=(r\sin\theta,r\cos\theta)$. 
The gas pressure $p_{\rm gas}$ inside the torus
is replaced with $p_{\rm gas}+p_{\rm rad}$
by assuming a local thermodynamic equilibrium, $T_{\rm g}=T_{\rm r}$.
The inner edge of the torus and 
the maximum pressure radius are 
set to $210~{\rm km}$ and $304.5~{\rm km}$, respectively.
The poloidal-loop magnetic fields inside the torus
are considered in addition to the dipole magnetic field of the NS.
The vector potential of this loop magnetic field 
is proportional to $\rho$.
The embedded loop magnetic field is anti-parallel 
to the dipole magnetic field at the inner edge of the torus
\citep{Romanova2011,Takahashi2017,Parfrey2017}.
We set the maximum $(p_{\rm gas}+p_{\rm rad})/p_{\rm mag}$ to 100 inside the torus.
We give a perturbation on $p_{\rm gas}+p_{\rm rad}$ by 10\% to 
break an equilibrium state of the torus.

The NS and torus are initially 
surrounded by relatively hot and low-density corona,
\begin{eqnarray*}
   \rho&=&
   \max\left[\rho_{\rm flr},
   \rho_{\rm c}\left(\frac{r_{\rm NS}}{r}\right)^{-5}\right],\\
   p_{\rm gas}&=&
   \max\left[p_{\rm flr},
   \min\left[\frac{k\rho T_{\rm g,max}}{\mu_{\rm w}m_{\rm p}},
   \frac{k\rho_c T_{\rm init}}{\mu_{\rm w}m_{\rm p}}
   \left(\frac{r_{\rm NS}}{r}\right)^{-5} \right]\right],
\end{eqnarray*}
where $\rho_{\rm c}$ is determined by
$\sigma_{\rm c}=B_{\rm NS}^2/(4\pi\rho_{\rm c})=10^3$,
and $T_{\rm init}$ is set to $5\times10^{11}~{\rm K}$.
The floor value of the gas density $\rho_{\rm flr}$ 
and gas pressure $p_{\rm flr}$
are described in Section \ref{sec:boundary}.
The radiation energy density of the corona 
is $\bar{E}=a\bar{T}_{\rm rad}^4$, 
where $\bar{T}_{\rm rad}=10^5~{\rm K}$.
The velocity of the gas $u^i$ and radiation rest frame $u_{\rm R}^i$
are set to zero in the corona.
The present result is not affected by the corona gas 
which is set initially. 
This is because these hot gases are blown away by 
the radiation force from the accretion disk 
after some time has elapsed since the simulations began.

{\subsection{Boundary and floor condition}\label{sec:boundary}}
\label{sec:boundray}

We adopt the outgoing boundary at $r=r_{\rm out}$, 
and the reflective boundary at $\theta=0,\pi$. 
At the inner boundary, $r=r_{\rm NS}$, 
we assume that the boundary, at which 
the gas can be swallowed by the NS 
but the energy can not be swallowed by the NS \citep{Ohsuga2007}.
At inner ghost cells, 
$\rho$, $p_{\rm gas}$, and $\bar{E}$ are the free boundary, 
and $u^i$ and $u^i_{\rm R}$ are set to zero.
For the magnetic field, 
the radial component is fixed to be the dipole field \citep{Wasserman1983},
while the tangential component 
is set according to the free boundary condition
\citep{Parfrey2017,Abarca2021}.
The numerical flux in mass conservation law,
energy conservation law for MHD, and
energy conservation law for the radiation field 
are zero at the NS surface.
The numerical flux of the induction equation
is also set to zero at the NS surface to satisfy the ideal MHD condition.
We also impose the following condition on 
the gas density and pressure at the NS surface:
\begin{eqnarray}
   \rho(t)&=&\min\left[\rho_*(t),~\rho(t=0)\right]
   \label{eq:inner_boundary1},\\
   p_{\rm gas}(t)&=&\frac{k\rho(t)T_{\rm g,*}(t)}{\mu m_{\rm p}},
   \label{eq:inner_boundary2}
\end{eqnarray}
where $\rho_*$ and $T_{\rm g,*}$ are the gas density and gas temperature 
calculated from the conservative variables, respectively.
In this case, the gas density is set to initial value 
by keeping the gas temperature constant at the NS surface.
The amount of reduced kinetic and thermal energy densities 
by applying
(\ref{eq:inner_boundary1}) and (\ref{eq:inner_boundary2}) is added to
the radiation energy density $R^t_t$.
Here, the kinetic and thermal energy densities 
are defined as follows:
\begin{eqnarray}
   U_{\rm kinetic}&=&\rho u^t(u_t-\sqrt{-g_{\rm tt}}),\\
   U_{\rm thermal}&=&(e+p_{\rm gas})u^t u_t+p_{\rm gas}g^t_t.
\end{eqnarray}
The mass reduced by the treatment of (\ref{eq:inner_boundary1}) is
considered to be swallowed by the NSs.

We impose a floor condition on $\rho$ and $p_{\rm gas}$ to
solve the GR-RMHD equations stably:
\begin{eqnarray}
   \rho_{\rm flr}&=&\max\left[
   \frac{b^2}{\sigma_{\rm max} },~
   10^{-4}\rho_0\left(\frac{r}{r_{\rm g}}\right)^{-1.5}
   \right],
   \label{eq:roflr}\\
   p_{\rm flr}&=&10^{-6}\left(\Gamma-1\right)\rho_0
   \left(\frac{r}{r_{\rm g}}\right)^{-2.5},
   \label{eq:eflr}
\end{eqnarray}
where $\sigma_{\rm max}$ is set to $10^3$.
The upper and lower limit of the gas temperature 
are set to
$5\times10^{11}~{\rm K}$ and
$5\times10^5~{\rm K}$, respectively.

The gas density should be low near $\theta=0,\pi$ 
because the strong magnetic pressure and tension 
due to the open poloidal magnetic field lines inhibit 
the mass flow toward the rotation axis 
in the very vicinity of the rotation axis.
In the numerical simulation, however, 
the gas reaches the region of $\theta=0,\pi$
due to the numerical diffusion. 
Then, the unphysical mass outflow driven by the radiation force 
is formed close to the polar axis.
We decided to ignore the gas-radiation interaction 
to avoid this problem in this region. 
Practically, we set the absorption and 
scattering opacity to zero at $\sigma > \sigma_{\rm rad}$, 
where $\sigma = b^2/ \rho$ is the magnetization.
In this study,
we set $\sigma_{\rm rad}=10$ in all models.
This value is set so that 
the opacity near the rotation axis is zero 
while the radiation from the dense accretion flows 
at the NS surface can be solved.

{\section{Results} \label{sec:results}}

In all models, 
the initial torus deviates from the equilibrium state
due to the magnetorotational instability (MRI)
caused by the differential rotation in the torus 
after the simulation starts.
The angular momentum of the gas is transported outward, 
and then the accretion disk is formed around the equatorial plane ($z=0$ plane).
The radiatively driven outflows are launched from the accretion disk.
The accretion disk in models 
{\tt\string B3e9D001}, {\tt\string B1e10D001}, 
{\tt\string B1e10D002}, {\tt\string B1e10D004}, {\tt\string B1e10D01}, 
{\tt\string B3e10D001}, {\tt\string B3e10D01}, {\tt\string B3e10D02}, 
{\tt\string B3e10D04} and {\tt\string B3e10D1}
is truncated before it reaches to the NS surface.
Then, the accreting gas moves along magnetic field lines, 
and column accretion flows (accretion columns) 
are formed near the poles of the NS.
On the other hand, the accretion disk in models 
{\tt\string B3e9D01\_np} and {\tt\string B1e10D1\_np}
directly connects to the NS surface without being truncated.

{\subsection{Quasi-steady state structure}\label{sec:quasi}}

\begin{figure*}[ht!]
\centering
\includegraphics[width=180mm]{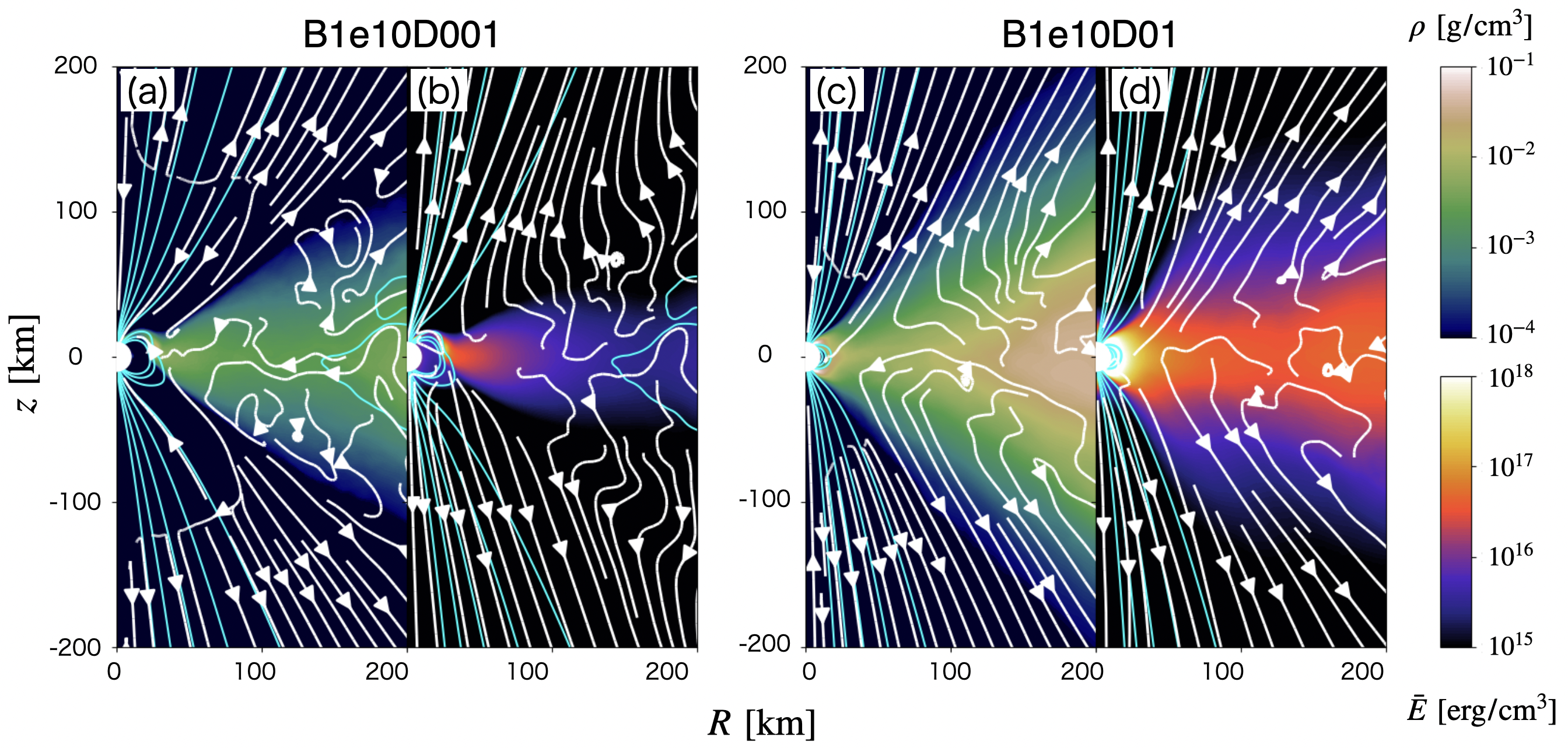}
\caption{
Here, we show the time-averaged quasi-steady structure 
of super-Eddington accretion flows 
around the magnetized NS 
for models {\tt\string B1e10D001} and {\tt\string B1e10D01}. 
Panels (a) and (c): Plot of density profile along with streamlines.
Panels (b) and (d): Colors show radiation energy density 
in the radiation rest frame 
and streamlines indicate the radiative flux.
Cyan lines are magnetic field lines.
\label{fig:figure2}}
\end{figure*}

\begin{figure*}[ht!]
\centering
\includegraphics[width=180mm]{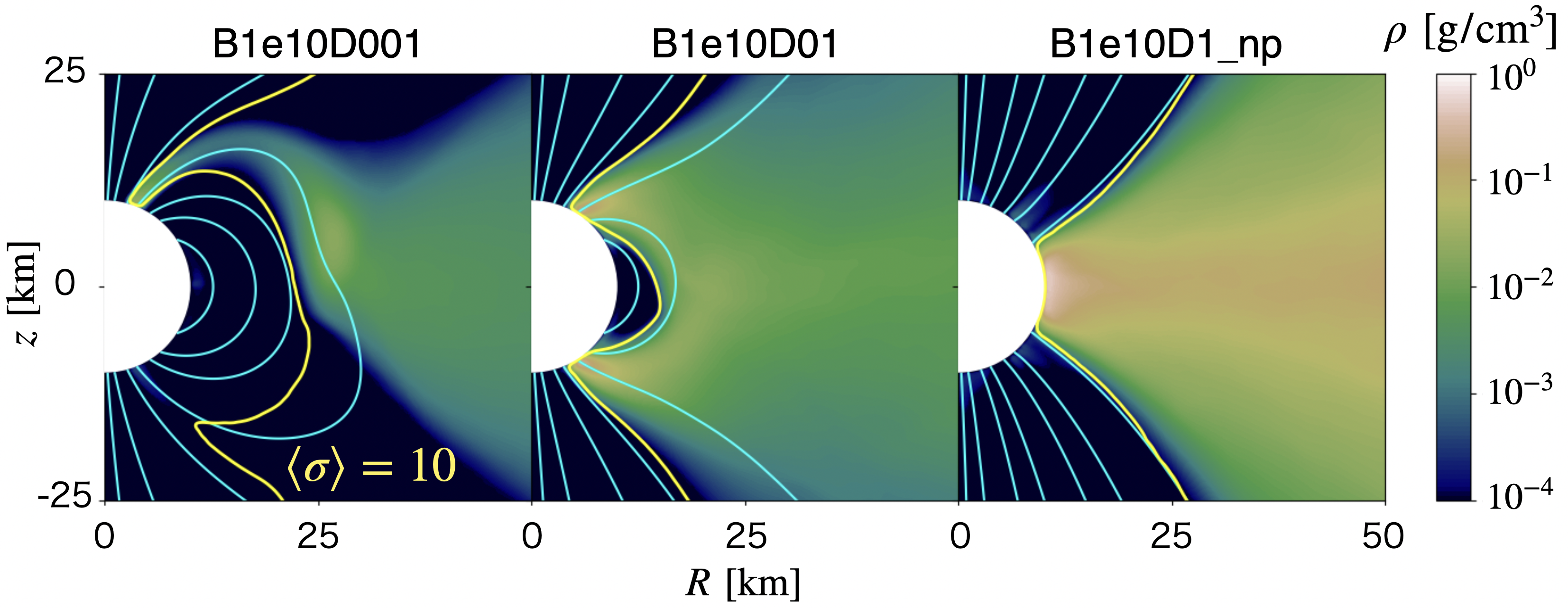}
\caption{
The dependence of accretion flows on 
the maximum gas density of the initial torus $\rho_0$.
Colors shows gas density.
Yellow lines represents $\left<\sigma\right>=10$, 
where $\left<\sigma\right>$ is the time-averaged magnetization.
Cyan lines are magnetic field lines.
\label{fig:figure3}}
\end{figure*}

We use the numerical data from period $t_{\rm eq}$ to $t_{\rm end}$ 
shown in Table \ref{tab:table1}
for investigating a quasi-steady structure in the present work, 
since the flows become quasi-steady and 
the mass outflow rate is almost constant after $t_{\rm eq}$.
Firstly, we introduce the quasi-steady 
structure of the accretion flows and outflows 
around the magnetized NS.
Figure \ref{fig:figure2} shows the time-averaged overview of 
models {\tt\string B1e10D001} (left) and {\tt\string B1e10D01} (right) 
in the quasi-steady state. 
Colors represent $\rho$ in panels (a) and (c),
and $\bar{E}$ in panels (b) and (d).
The white region with $r<10~{\rm km}$ is the NS,
and its center is located at the origin, $(R,z)=(0,0)$.
The magnetic field lines are shown by cyan solid lines.
We find that the accretion disk is formed 
near the equatorial plane at 
$20~{\rm km} \lesssim R \lesssim 200~{\rm km}$
(the green region in panels (a) and (c)).
In panels (b) and (d), it is clear that 
$\bar{E}$ is large in the accretion disk.
The accretion column is formed 
near the NS magnetic poles, $r \lesssim 20 {\rm km}$, 
the details of which are described in Figure \ref{fig:figure3}.
Streamlines follow the poloidal velocity vectors 
in panels (a) and (c).
It can be seen that the outflows are 
emanated from the disk toward the low-density region.
In addition, 
turbulent motions exist inside the accretion disk.
The radiative flux is represented by streamlines
in panels (b) and (d).
We can see that the outward 
radiative flux from the accretion disk appears.
The accretion disk, the outflows, and the outward radiative flux 
from the accretion disk are common to all models.
Both $\rho$ and $\bar{E}$ in the accretion disk tend to increase 
as $\rho_0$ becomes large.
Actually, both of $\rho$ and $\bar{E}$ 
in model {\tt\string B1e10D01} ($\rho_0=0.1~{\rm g~cm^{-3}}$)
are an order of magnitude higher than those 
in model {\tt\string B1e10D001} ($\rho_0=0.01~{\rm g~cm^{-3}}$).

We can see that 
the gas accretes onto the NS
around the pole outside the accretion column
($r\lesssim 100 {\rm km}$ and 
$\theta\lesssim40^\circ$, $140^\circ\lesssim\theta$ 
in panel (a)).
This is because the interaction between the gas and radiation 
is artificially switched off at the high-$\sigma$ region 
($\sigma > \sigma_\mathrm{rad}$) in our simulation 
to maintain numerical stability.
The radiation force does not work 
around the polar axis and the gas falls down to the NS. 
We note that the density in this region is too small 
to contribute to the mass accretion rate, 
and the effects on the overall structure 
are negligibly small.

Next, 
we explain the dependence of the magnetospheric structure on $\rho_0$.
Colors in Figure \ref{fig:figure3} show $\rho$, 
and magnetic field lines are represented by cyan lines.
We take models 
{\tt\string B1e10D001}, {\tt\string B1e10D01}, 
and {\tt\string B1e10D1\_np}.
We plot the time-averaged magnetization,
$\left<\sigma\right>={\left<b^2\right>}/{\left<\rho\right>}=10$,
by yellow contours.
Here, the angle brackets denote the time-averaged value.
The magnetization $\left<\sigma\right>$ on the high-density side 
of the yellow line is less than $10$.
In the left panel (model {\tt\string B1e10D001}), 
the high-density accretion column,
around $(R, z) = (5 {\rm km}, 12 {\rm km})$,
is formed (one-side accretion). 
Although not shown in Figure \ref{fig:figure3}, 
the same is true of 
models {\tt\string B3e9D001}, {\tt\string B1e10D002}, 
{\tt\string B3e10D001}, {\tt\string B3e10D01}, 
and {\tt\string B3e10D02}.
In the middle panel (model {\tt\string B1e10D01}), 
the accretion columns are formed 
at $(R, z) = (6 {\rm km}, \pm10 {\rm km})$,
which is true of models 
{\tt\string B1e10D004}, {\tt\string B3e10D04}, and {\tt\string B3e10D1}. 
At the truncation radius, 
which is the radius of the inner edge of the accretion disk,
the gas begins to move away from the equatorial plane toward the NS poles.
The truncation radius is $R\sim25~{\rm km}$ in model {\tt\string B1e10D001} 
and $R\sim13~{\rm km}$ in model {\tt\string B1e10D001},
and approximately corresponds to the position 
where $\left<\sigma\right>=10$ at the equatorial plane.
The density inside this radius
is less than 1\% of the density of the disk, 
except inside the accretion column. 
Also, the truncation radius is roughly comparable to
the magnetospheric radius, $r_{\rm M}$,
$\sim27~{\rm km}$ for model {\tt\string B1e10D001} 
and $\sim 13~{\rm km}$ for model {\tt\string B1e10D01}.
In the present work, 
$r_{\rm M}$ is defined as 
the maximum radius of the region, where $\left<\beta\right>_{u_\phi}<1$,
inside the maximum pressure radius of the initial torus, $304.5~{\rm km}$.
Here, $\left<\beta\right>_{u_\phi}$ is $u_\phi$-weighted plasma beta,
\begin{eqnarray}
    \left<\beta\right>_{u_\phi}=
    \frac{1}{\int^\pi_0 \left<u_\phi\right> 
    \sqrt{g_{\theta\theta}}d\theta}\int^\pi_0
    \frac{\left<p_{\rm gas}\right>+\left<p_{\rm rad}\right>}
    {\left<p_{\rm mag}\right>}
    \left<u_\phi\right> \sqrt{g_{\theta\theta}}d\theta.
    \label{eq:magneto_radius1}
\end{eqnarray}
Although $p_{\rm rad}$ is 
the radiation pressure in the radiation rest frame,
it is approximately equal to 
the radiation pressure in the fluid frame
since the MHD and radiation field 
inside the accretion disk are 
in local thermal equilibrium,
and therefore the radiation rest frame
coincides with the fluid frame.
We note that the gas pressure is negligible 
in the super-Eddington accretion disk \citep{Takahashi2017},
so that $p_{\rm mag}\sim p_{\rm rad}$ at $\left<\beta\right>_{u_\phi}=1$.
The coincidence between the truncation radius and $r_{\rm M}$ 
indicates that
the truncation radius is determined by the balance 
between $p_{\rm mag}$ and $p_{\rm rad}$.
As can be seen from the left and middle panels 
in Figure \ref{fig:figure3}, 
the truncation radius decreases as $\rho_0$ increases.

In model {\tt\string B1e10D1\_np}, 
since the radiation pressure of the accretion disk 
is larger than the magnetic pressure of the dipole field 
even at the NS surface, 
the accretion disk reaches the NS surface 
without being truncated and no accretion column is formed.
In this case, the region where $\left<\beta\right>_{u_\phi}>1$ 
reaches the NS surface, and 
$r_{\rm M}$ can not be determined.
Although not shown in this figure, 
accretion columns are not formed in model {\tt\string B3e9D01\_np}
for the same reason.
Based on the hypothesis that the radiation from the accretion column 
is the origin of the X-ray pulse, 
models {\tt\string B3e9D01\_np} and {\tt\string B1e10D1\_np} 
are inappropriate for explaining ULXPs.
Incidentally, the reason why the one-side accretion occurs
in model {\tt\string B1e10D001}
is described in Section \ref{sec:one-side}.

\begin{deluxetable*}{lcccccc}[tb]
\tablenum{2}
\tablewidth{0pt}
\tablehead{
\colhead{}
&\colhead{$\left<\right.\dot{M}_{\rm in}(r=11~{\rm km})\left.\right>$}
&\colhead{$\left<\right.\dot{M}_{\rm out}(r_{\rm out})\left.\right>$}
&\colhead{$\left<L_{\rm rad}(r_{\rm out})\right>$}
&\colhead{$\left<L_{\rm kin}(r_{\rm out})\right>$}
&\colhead{$T_{\rm bb}$}
&\colhead{$r_{\rm bb}$}
\\
\colhead{Model}
&\colhead{[$\dot{M}_{\rm Edd}$]}
&\colhead{[$\dot{M}_{\rm Edd}$]}
&\colhead{[$L_{\rm Edd}$]}
&\colhead{[$L_{\rm Edd}$]}
&\colhead{[$10^{7}$ K]}
&\colhead{[km]}
}
\startdata
{\tt\string B3e9D001}    & 32   & 36   & 3.0  & 0.15 & 0.72 & 57 \\
{\tt\string B3e9D01\_np} & 220  & 2700 & 130  & 47   & 0.63 & 230\\
{\tt\string B1e10D001}   & 54   & 74   & 4.1  & 0.34 & 0.71 & 58 \\
{\tt\string B1e10D002}   & 110  & 110  & 9.5  & 1.0  & 0.73 & 65 \\
{\tt\string B1e10D004}   & 160  & 730  & 21   & 7.1  & 0.66 & 130\\
{\tt\string B1e10D01}    & 500  & 2600 & 42   & 36   & 0.66 & 240\\
{\tt\string B1e10D1\_np} & 4700 & 23000& 500  & 410  & 0.50 & 1100\\
{\tt\string B3e10D001}   & 98   & 170  & 5.5  & 0.89 & 0.75 & 50 \\
{\tt\string B3e10D01}    & 720  & 2900 & 140  & 56   & 0.69 & 270\\
{\tt\string B3e10D02}    & 650  & 4600 & 150  & 37   & 0.63 & 250\\
{\tt\string B3e10D04}    & 1500 & 13000& 320  & 210  & 0.59 & 660\\
{\tt\string B3e10D1}     & 2300 & 28000& 490  & 460  & 0.55 & 1300
\enddata
\caption{
Time-averaged value for the mass accretion rate, 
mass outflow rate, radiative luminosity, and kinetic luminosity.
The mass accretion rate is measured at $r=11~\rm km$, 
and the others are calculated at $r=r_{\rm out}$.
\label{tab:table2}
}
\end{deluxetable*}

Table \ref{tab:table2} lists the time-averaged value of
the mass accretion rate $\dot{M}_{\rm in}$ at $r=11~\rm km$, 
mass outflow rate $\dot{M}_{\rm out}$ at $r=r_{\rm out}$, 
radiative luminosity $L_{\rm rad}$ at $r=r_{\rm out}$, and 
kinetic luminosity $L_{\rm kin}$ at $r=r_{\rm out}$ of each model. 
The Eddington mass accretion rate is denoted by $\dot{M}_{\rm Edd}=L_{\rm Edd}/c^2$.
The mass accretion rate and outflow rate are defined as
\begin{eqnarray}
    \dot{M}_{\rm in}(r)&=&-\int\min[\rho u^r,0]\sqrt{-g}d\theta d\phi,\\
    \dot{M}_{\rm out}(r)&=&\int\max[\rho u^r,0]\sqrt{-g}d\theta d\phi.
\end{eqnarray}
The raditive and kinetic luminosity can be calculated 
using following formula \citep{Sadowski2016}:
\begin{eqnarray}
    L_{\rm rad}(r)&=&-\int\min[R^r_t,0]
    \sqrt{-g}d\theta d\phi,\\
    L_{\rm kin}(r)&=&-\int\min[\rho u^r \left(u_t+\sqrt{-g_{tt}}\right),0]
    \sqrt{-g}d\theta d\phi.
    \label{eq:kinetic_luminosity}
\end{eqnarray}
As can be seen from this table,
$\dot{M}_{\rm in}$, $\dot{M}_{\rm out}$, $L_{\rm rad}$ and $L_{\rm kin}$ 
tend to be large as $\rho_0$ increases.
We also find that 
$L_{\rm kin}/L_{\rm rad}$ tends to be large as $\dot{M}_{\rm in}$ increases.
Indeed, the mass accretion rate is $54\dot{M}_{\rm Edd}$ and
$L_{\rm kin}/L_{\rm rad}\sim0.083$ in the case of model {\tt\string B1e10D001},
mass accretion rate is $500\dot{M}_{\rm Edd}$ and
$L_{\rm kin}/L_{\rm rad}\sim0.86$ in the case of model {\tt\string  B1e10D01},
and mass accretion rate is $4700\dot{M}_{\rm in}$ and
$L_{\rm kin}/L_{\rm rad}\sim1.2$ in the case of model  {\tt\string  B1e10D1\_np}.
That is, the kinetic luminosity is comparable to 
the radiation luminosity for the model with high mass accretion rate.
These trends are also reported 
in non-relativistic radiation hydrodynamics simulation 
\citep{Ohsuga2007}.

\begin{figure*}[ht!]
\centering
\includegraphics[width=180mm]{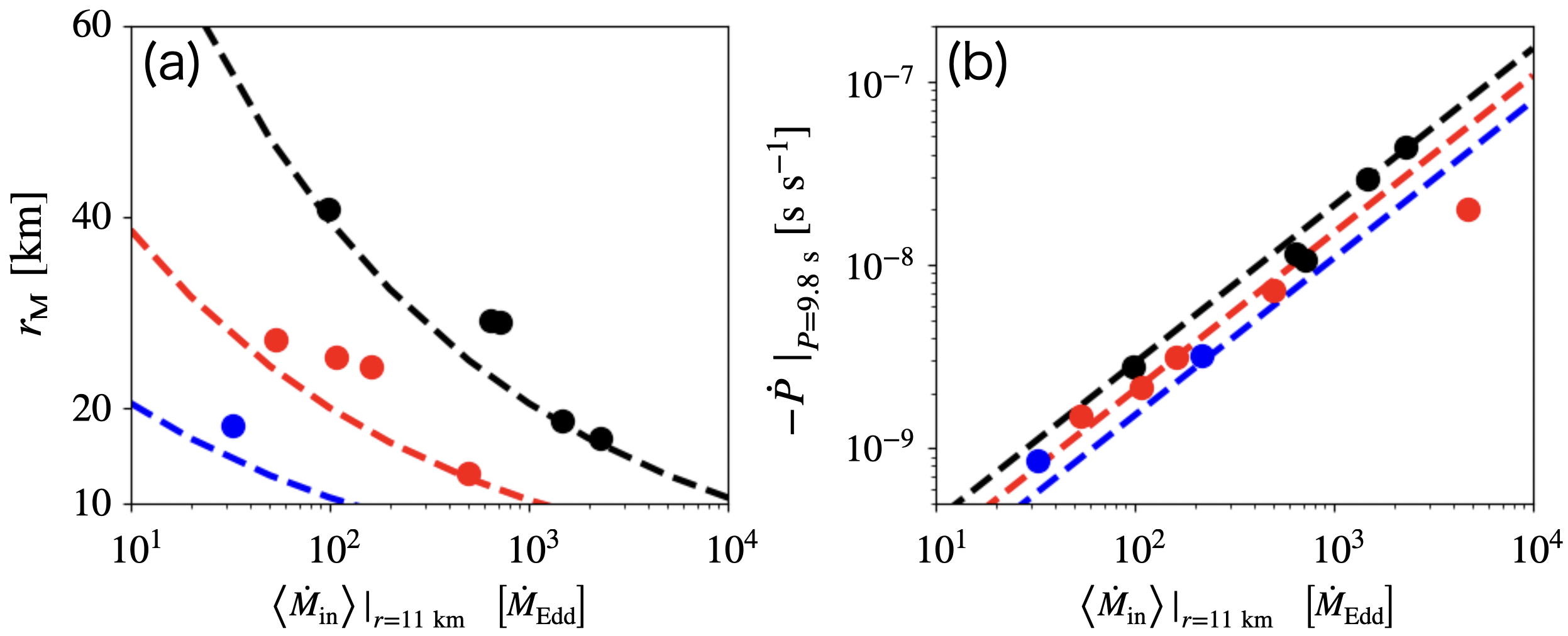}
\caption{
Panel (a): The relations between $r_{\rm M}$ and 
the mass accretion rate at the NS surface
with different surface magnetic field strengths,
$B_{\rm NS}= 3.3\times10^{9}~{\rm G}$ (red), 
$10^{10}~{\rm G}$ (blue),
and $3.3\times10^{10}~{\rm G}$ (black).
Dashed lines show the analytical model of $r_{\rm M}$ \citep{Takahashi2017}.
Panel (b): $\dot{P}$ as a function of 
the mass accretion rate at the NS surface
with different surface magnetic field strengths.
Dashed lines show the analytical model of $\dot{P}$ \citep{Takahashi2017}.
Parameters for the analytical model : 
$M_{\rm NS}=1.4M_\odot$, $r_{\rm NS}=10~{\rm km} $ and $\alpha=0.1$.
\label{fig:figure4}}
\end{figure*}

Figure \ref{fig:figure4} shows 
$r_{\rm M}$ and the time derivative 
of the NS rotation period (spin-up rate) $\dot{P}$
estimated using our numerical models 
as a function of the mass accretion rate.
Points in Figure \ref{fig:figure4} (a) indicate the resulting $r_{\rm M}$
for $B_{\rm NS}= 3.3\times10^{9}~{\rm G}$ (red), 
$10^{10}~{\rm G}$ (blue), 
and $3.3\times10^{10}~{\rm G}$ (black),
in which all models have accretion columns.
The magnetospheric radius $r_{\rm M}$
decreases with increasing $\dot{M}_{\rm in}$ 
and increases with increasing $B_{\rm NS}$.
Dashed lines indicate the analytical solution for $r_{\rm M}$, 
which can be calculated assuming that the magnetic pressure of the dipole field
balances with the radiation pressure of the accretion disk 
\citep[][see also Appendix \ref{sec:appendix2} for detail]{Takahashi2017}:
\begin{eqnarray}
r_{\rm M}&\sim& 2.0 \times10^6~[\rm cm]\nonumber\\
&\times&
\left(\frac{\alpha}{0.1}\right)^{2/7}
\left(\frac{\dot{M}_{\rm in}}{10^2\dot{M}_{\rm Edd}}\right)^{-2/7}
\left(\frac{B_{\rm NS}}{10^{10}~{\rm G}}\right)^{4/7}\nonumber\\
&\times&
\left(\frac{M_{\rm NS}}{1.4M_\odot}\right)^{-3/7}
\left(\frac{r_{\rm NS}}{10~{\rm km}}\right)^{12/7},
\label{eq:rM_analize}
\end{eqnarray}
where $\alpha$ is the viscous parameter \citep{Shakura1973}
and we take $\alpha=0.1$.
We can find that the resulting $r_{\rm M}$
is consistent with the analytical one.
We note that the results of models {\tt\string B1e10D1\_np} 
and {\tt\string B3e9D01\_np} are not plotted since $r_{\rm M}$
is not determined in these models as mentioned above.

We estimate $\dot{P}$ as follows,
\begin{eqnarray}
    \dot{P}&=&\frac{\left<\dot{L}\right>}{M_{\rm NS}l_{\rm NS}}P,
    \label{eq:spin-up-rate}\\
    \dot{L}&=&\int M^r_\phi \sqrt{-g}d\theta d\phi.
\end{eqnarray}
\citep{Shapiro1986,Takahashi2017}.
Here, $l_{\rm NS}=2\pi r_{\rm NS}^2/P$ and $P$
are the specific angular momentum and 
rotation period of the NS, respectively.
For numerical estimation, 
we assume $P=9.8~{\rm s}$, 
which corresponds to that observed in Swift J0243.6+6124. 
Figure \ref{fig:figure4} (b) shows 
that $\dot{P}$ is in the range from 
$-10^{-9}~{\rm s~s^{-1}}$ to $-10^{-7}~{\rm s~s^{-1}}$,
which is not inconsistent with the observed value of Swift J0243.6+6124, 
$\dot{P}\sim-10^{-8}~{\rm s~s^{-1}}$ 
(see Section \ref{sec:outflow_Bfield} for detail).
It is also clearly seen that 
the larger $\dot{M}_{\rm in}$ and higher $B_{\rm NS}$ 
leads to large $\dot{P}$.
Dashed lines indicate the analytical solution for $\dot{P}$,
which can be calculated assuming that 
the Keplerian angular momentum at $r_{\rm M}$ 
is transported to the NS without dissipation \citep[][see also Appendix \ref{sec:appendix2}]{Takahashi2017}:
\begin{eqnarray}
\dot{P}&\sim& -2.2\times 10^{-9}~[\rm s\cdot s^{-1}]\nonumber\\
&\times&
\left(\frac{\alpha}{0.1}\right)^{1/7}
\left(\frac{\dot{M}_{\rm in}}{10^2\dot{M}_{\rm Edd}}\right)^{6/7}
\left(\frac{B_{\rm NS}}{10^{10}~{\rm G}}\right)^{2/7}\nonumber\\
&\times&
\left(\frac{M_{\rm NS}}{1.4M_\odot}\right)^{2/7}
\left(\frac{r_{\rm NS}}{10~{\rm km}}\right)^{-8/7}
\left(\frac{P}{10~{\rm s}}\right)^2.
\label{eq:Pdot_analize}
\end{eqnarray}
The resulting $\dot{P}$ is roughly comparable to the analytical one.

\subsection{Powerful outflows}
\label{sec:outflow}

\begin{figure}[tb!]
\centering
\includegraphics[width=85mm]{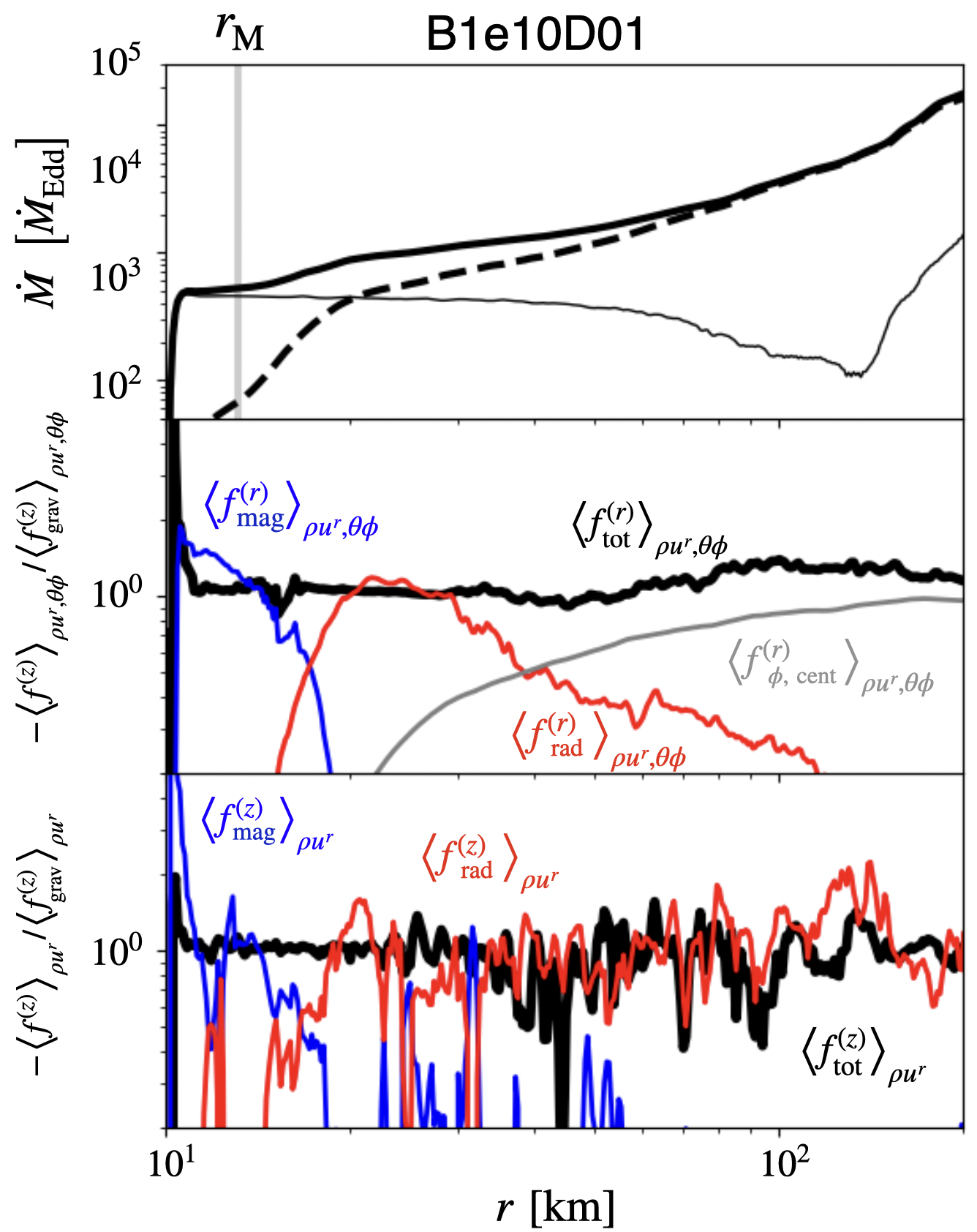}
\caption{
Top: 
time-averaged radial profiles of the mass accretion rate 
(thick solid lines), 
mass outflow rate (dashed lines),
and net flow rate (thin solid lines).
The vertical line is the position of the magnetospheric radius.
Middle: 
time and shell-averaged mass flux weighted 
radial forces normalized by the gravitational force.
Bottom: 
time-averaged mass flux weighted vertical forces 
divided by the vertical gravitational force at $\theta=\pi/2-\Theta_{\rm H}$.
\label{fig:figure5}}
\end{figure}

In Figure \ref{fig:figure5},
we show that 
powerful outflows are launched from the super-Eddington accretion disk 
and are driven by radiation and centrifugal forces.
The top panel in Figure \ref{fig:figure5} shows
time-averaged 
$\dot{M}_{\rm in}$ (thick solid line),
$\dot{M}_{\rm out}$ (dashed line), 
and
net flow rate $\dot{M}_{\rm net}=\dot{M}_{\rm in}-\dot{M}_{\rm out}$ 
(thin solid line) 
as a function of $r$ in the case of model {\tt\string B1e10D01}.
The position of $r_{\rm M}$ in this model
is represented by a vertical line.
This panel indicates that outflows mainly originate from 
the accretion disk.
This is clearly understood from the fact that 
$\dot{M}_{\rm out}$ increases significantly with increasing $r$,
and $\dot{M}_{\rm out}$ for $r<r_{\rm M}$ is much smaller than 
that for $r>r_{\rm M}$.
The mass accretion rate $\dot{M}_{\rm in}$ 
decreases with decreasing $r$ 
due to efficient mass ejection from the accretion disk.
The mass outflow rate $\dot{M}_{\rm out}$ is much smaller than 
the mass accretion rate at $r<r_{\rm M}$,
and we find $\dot{M}_{\rm net}\sim\dot{M}_{\rm in}\gg\dot{M}_{\rm out}$
inside the magnetosphere.

Next, we investigate the radial profile of forces acting on fluid elements. 
The steady-state axisymmetric equations of motion 
in the $r$- and $\theta$-direction in the statistic observer frame are,
\begin{eqnarray}
f_{\rm grav}^{(r)}
+f_{\rm thermal}^{(r)}
+f_{\rm \theta,cent}^{(r)}
+f_{\rm \phi,cent}^{(r)}
+f_{\rm rad}^{(r)}
+f_{\rm mag}^{(r)}
+f_{\rm adv}^{(r)}&=&0,\\
f_{\rm inertial}^{(\theta)}
+f_{\rm thermal}^{(\theta)}
+f_{\rm \phi,cent}^{(\theta)}
+f_{\rm rad}^{(\theta)}
+f_{\rm mag}^{(\theta)}
+f_{\rm adv}^{(\theta)}&=&0,
\end{eqnarray}
where superscript $(r)$ and $(\theta)$ denote 
the $r$- and $\theta$-component of the vectors 
in the static observer frame. 
The forces acting on fluid elements 
consist of the gravitational force $f_{\rm grav}$, 
thermal force $f_{\rm thermal}$, 
radiation force $f_{\rm rad}$, 
Lorentz force $f_{\rm mag}$, 
centrifugal force $f_{\rm cent}$, 
inertial force $f_{\rm inertial}$,
and advection force $f_{\rm adv}$. 
The centrifugal force is divided into two components, 
$f_{\theta, \rm cent}$ and $f_{\phi, \rm cent}$, 
which are 
caused by the polar ($\theta$-) and azimuthal ($\phi$-) motions,
respectively
(see Appendix \ref{sec:appendix3} for detail).
We define the total force $f_{\rm tot}$ as the sum of all forces 
without $f_{\rm grav}$,
and therefore $-f_{\rm tot}^{(r)}/f_{\rm grav}^{(r)}=1$ holds.

The middle panel in Figure \ref{fig:figure5} shows 
time and shell-averaged mass flux weighted radial forces 
normalized by the gravitational force
as a function of $r$,
which can be calculated as follows:
\begin{eqnarray}
    \left<f^{(r)}\right>_{\rho u^r,\theta\phi}
    =\frac{1}{\left<\right.\dot{M}_{\rm out}(r)\left.\right>}
    \left<\int f^{(r)}
    \max[\rho u^r,~0] \sqrt{-g}d\theta d\phi\right>. 
\end{eqnarray}
The bottom panel in Figure \ref{fig:figure5} shows 
time-averaged mass flux weighted $z$-component of the forces 
normalized by the vertical gravitational force,
\begin{eqnarray}
\left<f^{(z)}\right>_{\rho u^r} 
&=&\frac{\left<\max[\rho u^r,0]f^{(z)}\right>}{\left<\max[\rho u^r,0]\right>},\\
f^{(z)}&=&f^{(r)}\cos\theta-f^{(\theta)}\sin\theta,
\end{eqnarray}
at the disk surface.
The polar angle of the disk surface is defined 
as $\theta=\pi/2-\Theta_{\rm H}$, 
where
\begin{eqnarray}
\Theta_{\rm H}=
\left<\sqrt{\frac{\int\rho(\pi/2-\theta)^2\sqrt{g_{\rm \theta\theta}}d\theta}{\int\rho\sqrt{g_{\rm \theta\theta}}d\theta}}
\right>.
\end{eqnarray}
It is clear that the outflows are mainly driven 
by the radiation force for $20~{\rm km}<r<50~{\rm km}$,
where the radiation force is stronger than 
the other forces in both $r$- and $z$-directions.
For $50~{\rm km}<r<200~{\rm km}$, 
the gas is launched from the accretion disk by the radiation force
and moves outward by centrifugal force.
The Lorentz force is dominant at $r<r_{\rm M}$, 
but as already mentioned, 
most of the outflows are of disk origin ($r>r_{\rm M}$). 
The thermal force, 
centrifugal force due to the motion in the $\theta$-direction, 
advection force,
and inertial force are significantly small over the whole region.
The radial distribution of forces for $r>r_{\rm M}$ 
in models
{\tt\string B3e9D001}, 
{\tt\string B3e9D01}, 
{\tt\string B1e10D001},
{\tt\string B1e10D002},
{\tt\string B1e10D004},
{\tt\string B3e10D001},
{\tt\string B3e10D01},
{\tt\string B3e10D02},
{\tt\string B3e10D04},
and {\tt\string B3e10D1}
is similar to that in model {\tt\string B1e10D01}.
The radial profile of forces outside $r=r_{\rm NS}$ 
in models {\tt\string B3e9D01\_np} and {\tt\string B1e10D1\_np},
where the magnetosphere is not formed,
is the same as that in the disk region 
outside $r=r_{\rm M}$ in model {\tt\string B1e10D01}.

\begin{figure*}[ht!]
\centering
\includegraphics[width=180mm]{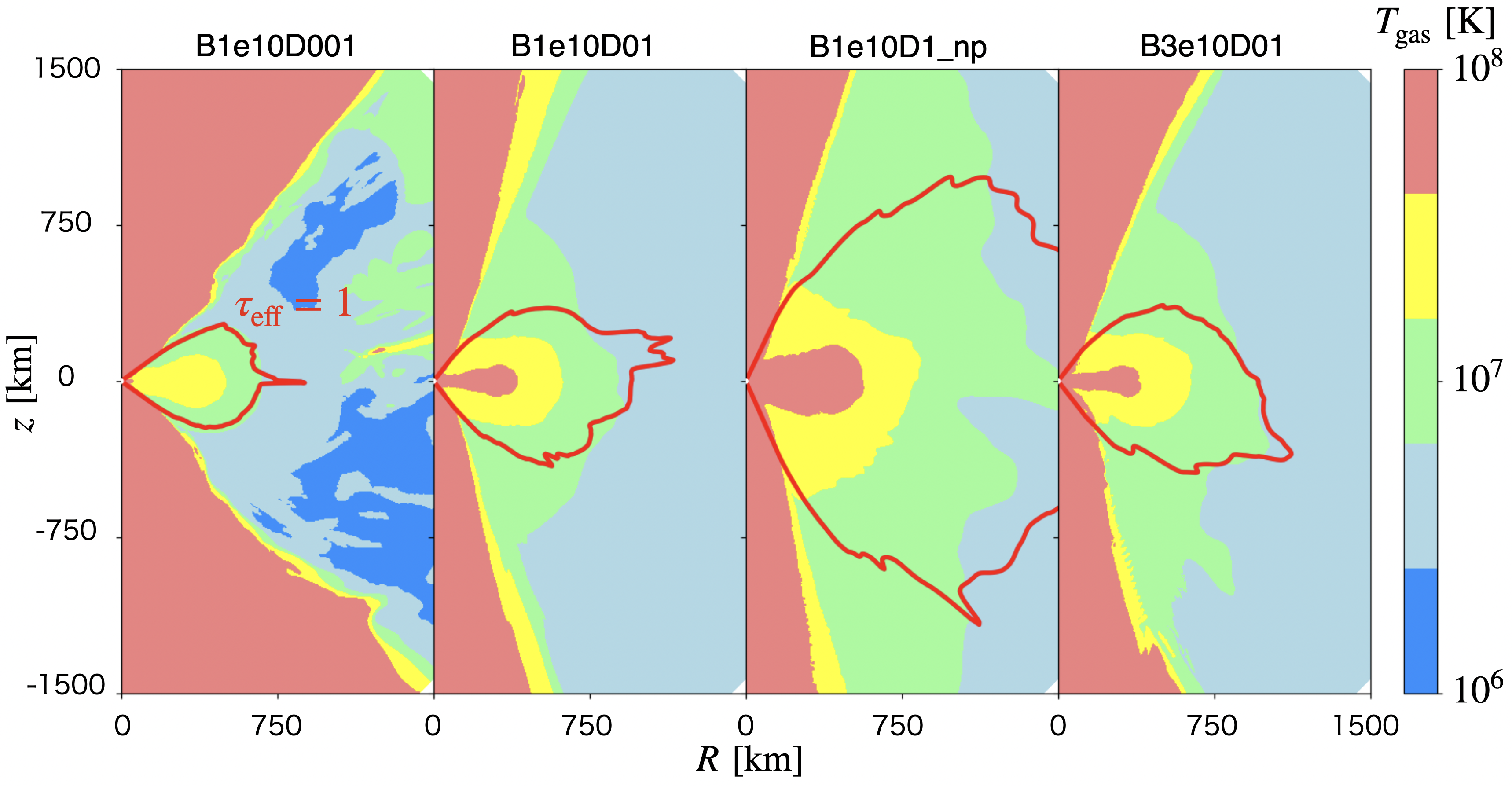}
\caption{
Distributions of time-averaged gas temperature of 
models 
{\tt\string B1e10D001}, 
{\tt\string B1e10D01}, 
{\tt\string B1e10D1\_np}, 
Red lines show locations of the photosphere, 
$\tau_{\rm eff}=1$.
\label{fig:figure6}}
\end{figure*}

In Figure \ref{fig:figure6}, 
we show the gas temperature distribution and 
the location of the photosphere.
In all models, 
the gas temperature tends to be high near the $z$-axis 
and in the equatorial plane at $r<750~{\rm km}$. 
Red lines indicate the photospheres, 
$\left<\tau_{\rm eff}\right>=1$, 
where $\left<\tau_{\rm eff}\right>$ 
is the time-averaged effective optical depth measured as,
\begin{eqnarray}
    \left<\tau_{\rm eff}(r)\right>
    =\int^{r_{\rm out}}_r \left<\rho\right> 
    \sqrt{\left<\kappa_{\rm abs}\right>
    (\left<\kappa_{\rm abs}\right>
    +\kappa_{\rm sca})}\sqrt{g_{rr}}dr,
    \label{eq:efftau}
\end{eqnarray}
where 
$\left<\kappa_{\rm abs}\right>
=6.4\times 10^{22}\left<\rho\right> \left<T_{\rm g}\right>^{-3.5}
~[{\rm cm^2~g^{-1}}]$.
As can be seen in this figure, 
the photosphere extends to surround the disk region. 
The larger the mass accretion rate, 
the wider the region surrounded by the photosphere. 
This can be understood from a comparison of 
models {\tt\string B1e10D001},
{\tt\string B1e10D01},
and {\tt\string B1e10D1\_np}. 
Also, within the range of parameters employed in the present study, 
the position of the photosphere does not depend much on $B_{\rm NS}$.
It is clear from the fact that the photosphere of 
models {\tt\string B1e10D01} and {\tt\string B3e10D01} 
hardly change (see red lines in Figure \ref{fig:figure6}).
In all models, the gas temperature on the photosphere 
is $\sim10^7~{\rm K}$ over a wide range 
of polar angles. 
From this, it is expected that blackbody radiation 
with a temperature of $\sim10^7~{\rm K}$ can be observed.

\begin{figure}[ht!]
\centering
\includegraphics[width=85mm]{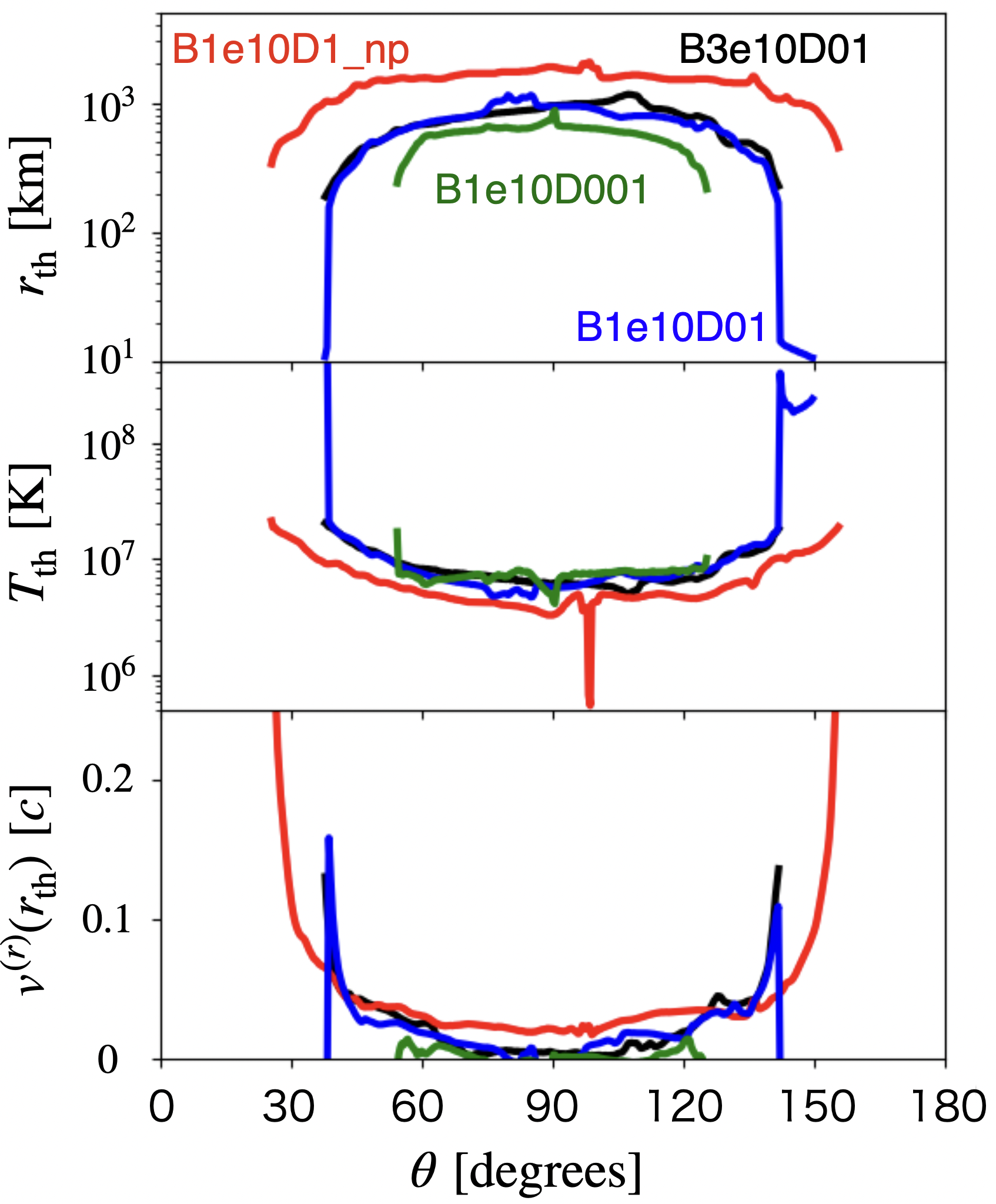}
\caption{
The polar angle ($\theta$) dependencies of 
the thermalization radius $r_{\rm th}$ (top panel), 
the gas temperature at the photospheres $T_{\rm th}$ (middle panel), and 
the gas velocity at the photospheres $v_{\rm th}$ (bottom panel).
\label{fig:figure7}}
\end{figure}

Hereafter, we define the thermalization radius $r_{\rm th}$ 
as the radius where $\left<\tau_{\rm eff}\right>=1$. 
That is, the radius of the photosphere is represented as $r_{\rm th}$.
In Figure \ref{fig:figure7}, 
we plot $r_{\rm th}$ (upper panel) 
and the time-averaged gas temperature at the photospheres 
$T_{\rm th}$ (middle panel)
as a function of polar angle $\theta$.
In model {\tt\string B1e10D1\_np}, 
which has the largest $\dot{M}_{\rm in}$ in our models,
$r_{\rm th}\gtrsim1000~{\rm km}$ at $30^\circ<\theta<150^\circ$.
Models with smaller $\dot{M}_{\rm in}$ 
tend to have smaller $r_{\rm th}$.
The thermalization radius $r_{\rm th}$
is slightly less than $1000~{\rm km}$ at $40^\circ<\theta<140^\circ$
in models {\tt\string B1e10D01} and {\tt\string B3e10D01}
and is $\sim500~{\rm km}$ at $60^\circ<\theta<120^\circ$
in model {\tt\string B1e10D001}.
The thermalization radius cannot 
be determined 
close to the rotation axis 
($\theta \lesssim 30^\circ$, $150^\circ\lesssim \theta$) 
since the gas is optically thin and 
$\left<\tau_{\rm eff}\right><1$ even at the NS surface. 
The range of $\theta$, 
within which $r_{\rm th}$ can be obtained, 
narrows as $\dot{M}_{\rm in}$ decreases.
In all models, 
$T_{\rm th}$ is almost constant at the angular range 
where $r_{\rm th}$ can be determined. 
The bottom panel in Figure \ref{fig:figure7} shows
the polar angle dependence of 
the radial gas velocity at the photosphere, $v^{(r)}(r_{\rm th})$,
where $v^{(r)}=u^{(r)}/u^{(t)}$.
We can see $v^{(r)}(r_{\rm th})>0$ on the photosphere,
which indicates that 
most of the photosphere is located 
inside the optically thick outflows.
Hereafter, we refer to such an outflow 
as an effectively optically thick outflow
\citep{Urquhart2016}.

Here, we estimate the blackbody temperature $T_{\rm bb}$ 
by taking $\theta$-averaged $T_{\rm th}$ on the photosphere as
\begin{eqnarray}
T_{\rm bb}=\frac{1}{\int \sqrt{-g(r_{\rm th}(\theta),\theta)}d\theta}
\int T_{\rm th}(\theta)
\sqrt{-g(r_{\rm th}(\theta),\theta)}d\theta.
\end{eqnarray}
This integration is performed on the surface 
where the thermalization radius can be estimated.
The corresponding blackbody radius $r_{\rm bb}$ can be calculated as 
\begin{eqnarray}
r_{\rm bb}
&=&
\left[\frac
{\left<L_{\rm ISO}\right>_\theta}
{4\pi\sigma  T_{\rm bb}^4}
\right]^{1/2},
\end{eqnarray}
where the $\theta$-averaged isotropic luminosity 
$\left<L_{\rm ISO}\right>_\theta$ is 
\begin{eqnarray}
\left<L_{\rm ISO}\right>_\theta
&=&
-\frac{4\pi r_{\rm out}^2 }{\int \sqrt{-g(r_{\rm out},\theta)}d\theta}
\int \left<R^r_t(r_{\rm out})\right>\sqrt{-g(r_{\rm out},\theta)}d\theta.
\end{eqnarray}
Here, $\theta$ integration is also conducted only in 
the range where the thermalization radius can be obtained.
The resulting $T_{\rm bb}$ and 
$r_{\rm bb}$ are presented in Table \ref{tab:table2}.
The blackbody temperature
is not so sensitive to the accretion rate
and is $\sim10^7~{\rm K}$,
which is consistent with the blackbody temperature
observed in Swift J0243.6+6124 at the super-Eddington phase
\citep{Tao2019}. 
The resulting $r_{\rm bb}$ is roughly fitted as 
\begin{eqnarray}
    r_{\rm bb}=3.2\left(\frac{\dot{M}}{\dot{M}_{\rm Edd}}\right)^{0.71}
    ~[\rm km].
    \label{eq:fitting}
\end{eqnarray}
\citet{Tao2019} showed that the thermal photons 
would come from $r=100-500~{\rm km}$ for Swift J0243.6+6124. 
Using equation (\ref{eq:fitting}), 
our model can explain this radius 
when the accretion rate is 
$130 \dot{M}_{\rm Edd} < \dot{M}_{\rm in} < 1200\dot{M}_{\rm Edd}$. 
In the models treated in the present study, 
there is no clear result that $r_{\rm bb}$ 
depends on $B_{\rm NS}$.
We will discuss this point later.


We run three models, 
{\tt\string B1e10D001\_a}, {\tt\string B1e10D001\_b}, 
and {\tt\string B1e10D001\_c}, 
which have the same initial parameters as 
model {\tt\string B1e10D001} except for the resolution,
to assess the effect of the resolution.
We confirm that $r_{\rm M}$ and $\dot{M}_{\rm out}$ of models
{\tt\string B1e10D001\_a} and model {\tt\string B1e10D001\_b}
are almost the same as those of model {\tt\string B1e10D001},
while those of model {\tt\string B1e10D001\_c} are different from 
those of model {\tt\string B1e10D001}.
The resolution of the simulation models adopted in this study 
is the same as or higher than that of model {\tt\string B1e10D001\_b}.
Therefore, our simulation results are independent of the resolution.

\subsection{Magnetic field strength of the neutron star in Swift J0243.6+6124\label{sec:outflow_Bfield}}

Here, we suggest how to constrain $B_{\rm NS}$ 
using analytical solutions of $r_{\rm M}$ and $\dot{P}$
and the conditions under which effectively optically thick outflows, 
which can reproduce the blackbody radiation observed in Swift J0243.6+6124,
occur.

It is necessary to simultaneously satisfy $r_{\rm M}<r_{\rm co}$, 
where $r_{\rm co}$ is the corotation radius,
and $r_{\rm M}$ can be defined 
(that is, $r_{\rm M}>r_{\rm NS}$) in ULXPs.
The condition of $r_{\rm M}<r_{\rm co}$ is 
for gas to accrete onto the NS \citep{Illarionov1975}.
Otherwise, the gas would all be blown away due to the propeller effect.
The accretion columns are formed in the case 
that the second condition is satisfied
(see Figure \ref{fig:figure3} for detail).
Substituting equations (\ref{eq:rM_analize}) and (\ref{eq:Pdot_analize}) 
into these condition, we get 
\begin{eqnarray}
B_{\rm NS}&<&1.6\times10^{14}\left[\rm G\right]\nonumber\\
&\times&
\left(\frac{\alpha}{0.1}\right)^{-1/2}
\left(\frac{\dot{P}}{-10^{-8}~{\rm s~s^{-1}}}\right)^{1/2}\nonumber\\
&\times&
\left(\frac{M_{\rm NS}}{1.4M_\odot}\right)
\left(\frac{r_{\rm NS}}{10^6~{\rm cm}}\right)^{-2},
\label{eq:propeller_condition_BNS_p_pdot}
\end{eqnarray}
and
\begin{eqnarray}
B_{\rm NS}&>&7.5\times10^9~\left[\rm G\right]\nonumber\\
&\times&
\left(\frac{\alpha}{0.1}\right)^{-1/2}
\left(\frac{\dot{P}}{-10^{-8}~{\rm s~s^{-1}}}\right)^{1/2}
\left(\frac{M_{\rm NS}}{1.4M_\odot}\right)^{1/2}\nonumber\\
&\times&
\left(\frac{r_{\rm NS}}{10^6~{\rm cm}}\right)^{-1/2}
\left(\frac{P}{10~{\rm s}}\right)^{-1}.
\label{eq:pulse_condition_BNS_p_pdot}
\end{eqnarray}
If the magnetic field strength at the NS surface satisfies 
the inequalities of (\ref{eq:propeller_condition_BNS_p_pdot}) and 
(\ref{eq:pulse_condition_BNS_p_pdot}), 
then the column accretion onto the magnetic poles of the NS appears,
and such objects can be the candidates for the ULXPs. 

Next, we consider the emergence of the outflows 
from the super-Eddington disks.
The radiatively driven outflows are thought to 
be launched in the slim disk region, 
which is inside the spherization radius (or trapping radius), 
$r_{\rm sph}$. 
This is discussed in \citet{Shakura1973} 
and revealed by the numerical simulations 
\citep{Kitaki2021,Yoshioka2022}. 
The spherization radius is roughly estimated as 
$r_{\rm sph}=(3/2)(\dot{M}_{\rm in}/\dot{M}_{\rm Edd})r_{\rm g}$.
Since the disk is truncated at $r=r_{\rm M}$, 
the ULXPs accompanying the outflows appear
if the condition of $r_{\rm M}<r_{\rm sph}$
is satisfied in addition to $r_{\rm NS}<r_{\rm M}<r_{\rm co}$
\citep{Mushtukov2019}.
Using equations (\ref{eq:rM_analize}) and (\ref{eq:Pdot_analize}),
$r_{\rm M}<r_{\rm sph}$ becomes
\begin{eqnarray}
B_{\rm NS}
&<&
1.5\times10^{12}~\left[\rm G\right]\nonumber\\
&\times&
\left(\frac{\alpha}{0.1}\right)^{-1/2}
\left(\frac{\dot{P}}{-10^{-8}~{\rm s~s^{-1}}}\right)^{3/2}\nonumber\\
&\times&
\left(\frac{M_{\rm NS}}{1.4M_\odot}\right)
\left(\frac{P}{10~\rm s}\right)^{-3}.
\label{eq:wind_condition_BNS_p_pdot}
\end{eqnarray}

Equations 
(\ref{eq:propeller_condition_BNS_p_pdot})-(\ref{eq:wind_condition_BNS_p_pdot}) 
are the general conditions to explain the ULXPs,
while we concentrate on the discussion about Swift J0243.6+6124 
in the following.
As described in Section \ref{sec:outflow},
the range of the mass accretion rate 
that can reproduce the blackbody radiation observed in this object is 
$130\dot{M}_{\rm Edd}<\dot{M}_{\rm in}<1200\dot{M}_{\rm Edd}$
from relation (\ref{eq:fitting}).
Using equation (\ref{eq:Pdot_analize}), 
this condition is transformed as
\begin{eqnarray}
B_{\rm NS}
&<&9.8\times10^{11}~[{\rm G}]\nonumber\\
&\times&
\left(\frac{\alpha}{0.1}\right)^{-1/2}
\left(\frac{\dot{P}}{-10^{-8}~{\rm s~s^{-1}}}\right)^{7/2}
\left(\frac{M_{\rm NS}}{1.4M_\odot}\right)^{-1}\nonumber\\
&\times&
\left(\frac{r_{\rm NS}}{10^6~{\rm cm}}\right)^4
\left(\frac{P}{10~{\rm s}}\right)^{-7},
\label{eq:eoto_cond1}\\
B_{\rm NS}
&>&1.3\times10^{9}~[{\rm G}]\nonumber\\
&\times&
\left(\frac{\alpha}{0.1}\right)^{-1/2}
\left(\frac{\dot{P}}{-10^{-8}~{\rm s~s^{-1}}}\right)^{7/2}
\left(\frac{M_{\rm NS}}{1.4M_\odot}\right)^{-1}\nonumber\\
&\times&
\left(\frac{r_{\rm NS}}{10^6~{\rm cm}}\right)^4
\left(\frac{P}{10~{\rm s}}\right)^{-7}.
\label{eq:eoto_cond2}
\end{eqnarray}
Among ULXPs with outflows, 
objects that meet the conditions 
(\ref{eq:wind_condition_BNS_p_pdot}) and (\ref{eq:eoto_cond1}) 
would have effectively optically thick outflows,
and the blackbody radiation with $r_{\rm bb}>100~{\rm km}$ 
would appear in the radiation spectrum. 

\begin{figure}[ht!]
\centering
\includegraphics[width=85mm]{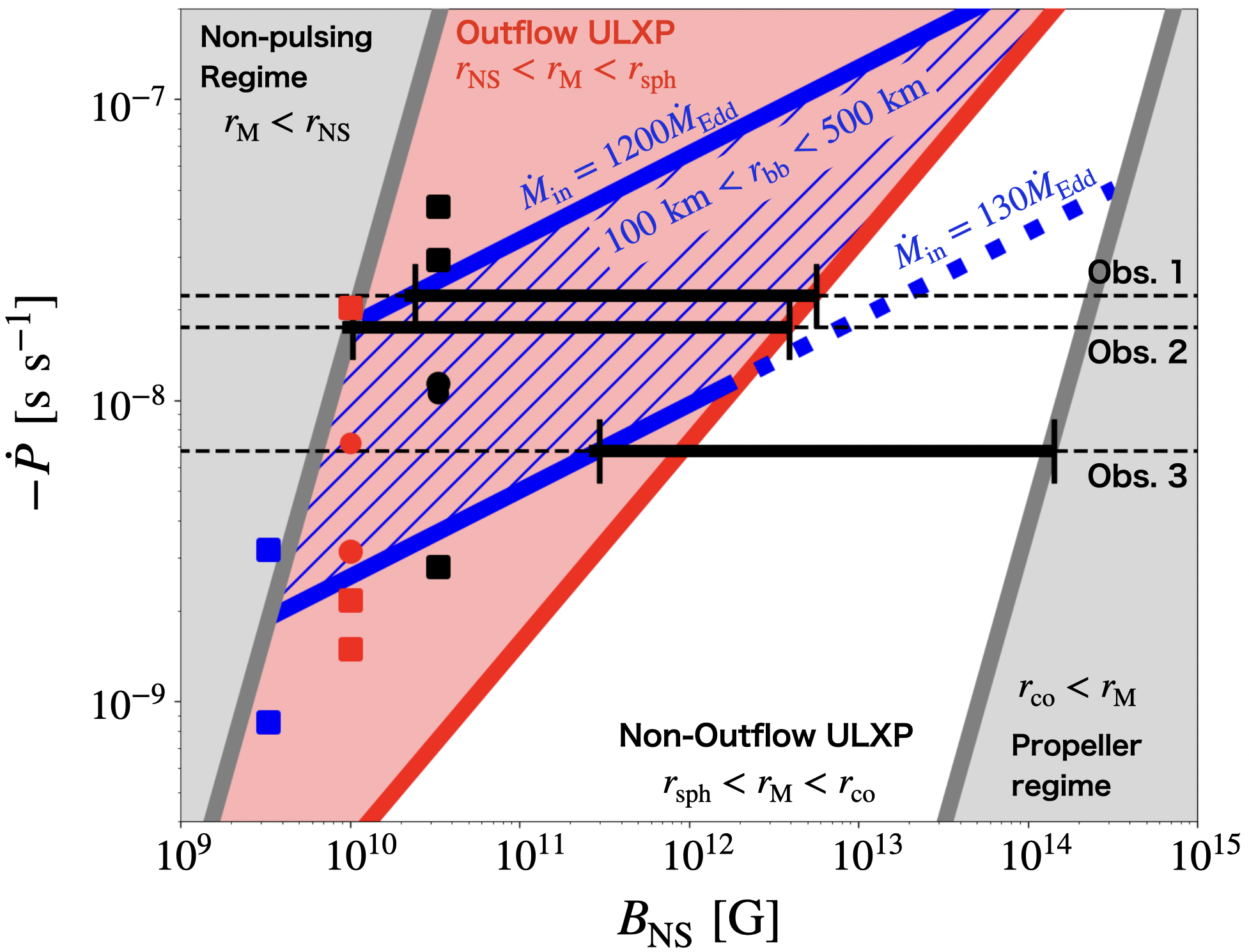}
\caption{
Here, the conditions 
(\ref{eq:propeller_condition_BNS_p_pdot})-(\ref{eq:eoto_cond2})
are summarized in the $\dot{P}-B_{\rm NS}$ plane (see text for detail).
The spin-up rates of 
models {\tt\string B1e10D004}, {\tt\string B1e10D01}, 
{\tt\string B3e10D01}, and {\tt\string B3e10D02}
are shown by filled circles,
while square markers are that of the other models.
We employ table 1 in \citet{Chen2021} for the observed $\dot{P}$ 
\citep[see also][]{Doroshenko2018}.
Parameters for the conditions:
$\alpha=0.1$,
$M_{\rm NS}=1.4M_\odot$, 
$r_{\rm NS}=10^6~{\rm cm}$, and $P=9.8~{\rm s}$.
\label{fig:figure8}}
\end{figure}

Figure \ref{fig:figure8} summarizes 
the conditions shown above in the $\dot{P}-B_{\rm NS}$ plane. 
Here, we assume $P=9.8~{\rm s}$, 
which is the rotation period observed in Swift J0243.6+6124 \citep{Doroshenko2018,Chen2021}.
We take $\alpha=0.1$, $M_{\rm NS}=1.4M_\odot$, and $r_{\rm NS}=10~{\rm km}$.
The red shaded region indicates the ULXP with outflow,
which satisfies both $r_{\rm NS}<r_{\rm M}<r_{\rm co}$ and $r_{\rm M}<r_{\rm sph}$.
The ULXP without outflow is in the white region, where 
$r_{\rm NS}<r_{\rm M}<r_{\rm co}$ is satisfied
but $r_{\rm M}<r_{\rm sph}$ is not satisfied.
The condition of ULXPs such as $r_{\rm M}>r_{\rm co}$ (propeller regime)
or $r_{\rm M}<r_{\rm NS}$ (non-pulsating regime) is not satisfied 
in grey shaded regions.
In the blue hatched region
where the accretion rate is 
$130\dot{M}_{\rm Edd}<\dot{M}_{\rm in}<1200\dot{M}_{\rm Edd}$,
effectively optically thick outflows 
with $r_{\rm bb}=100-500\rm km$
and $T_{\rm bb} \sim 10^7 \rm K$ appear so that 
the blackbody radiation, as observed in Swift J0243.6+6124, 
is thought to be reproduced.
Here, it is confirmed that 
the models 
that can successfully reproduce 
such effectively optically thick outflows and accretion columns
({\tt\string B1e10D004}, {\tt\string B1e10D01}, 
{\tt\string B3e10D01}, and {\tt\string B3e10D02})
are located in the blue hatched region 
(filled circles).
In contrast, other models are 
located outside the blue hatched region (filled squares).
In observations of Swift J0243.6+6124, $P$ is almost constant,
while $\dot{P}$ is reported to be 
$-2.22\times10^{-8}~{\rm s~s^{-1}}$ (Obs. 1),
$-1.75\times10^{-8}~{\rm s~s^{-1}}$ (Obs. 2), and
$-6.8 \times10^{-9}~{\rm s~s^{-1}}$ (Obs. 3)
\citep{Doroshenko2018,Chen2021}.
These observed spin-up rates are 
denoted by black dashed lines in Figure \ref{fig:figure8}.
In Obs. 1 and 2, the observed isotropic luminosity exceeds $L_{\rm Edd}$, 
and the thermal emission 
of which the observed blackbody radius is $100-500~{\rm km}$
is observed 
\citep[obsID: 90302319004, 90302319006 and 90302319008 in][]{Tao2019}.
Then,
the magnetic field strength is restricted to be
$2\times10^{10}~{\rm G}<B_{\rm NS}<5\times10^{12}~{\rm G}$ for Obs. 1 and $10^{10}~{\rm G}<B_{\rm NS}<4\times10^{12}~{\rm G}$ for Obs. 2,
which are represented by black solid lines.
In Obs. 3, on the other hand, 
the luminosity is lower than $L_{\rm Edd}$, 
and the thermal emission is not observed.
Such features are realized 
in the condition that the magnetic field is in the range of
$3\times10^{11}~{\rm G} < B_{\rm NS} < 10^{14}~{\rm G}$
(see black solid line for Obs. 3).
Therefore, the magnetic field strength 
that satisfies the three observations is between 
$3\times 10^{11}~{\rm G}$ and $4\times10^{12}~{\rm G}$.
Using the equation (\ref{eq:Pdot_analize}),
the mass accretion rate corresponding to this range
is estimated to be 
$200\dot{M}_{\rm Edd}\lesssim
\dot{M}_{\rm in}\lesssim500\dot{M}_{\rm Edd}$
for Obs. 1 and 2,
and 
$60\dot{M}_{\rm Edd}\lesssim
\dot{M}_{\rm in}\lesssim100\dot{M}_{\rm Edd}$
for Obs. 3.
Since the luminosity has been reported to be $\sim10^{39}~{\rm erg~s^{-1}}$ 
at the super-Eddington phase in Swift J0243.6+6124 (Obs.1 and 2), 
the radiative efficiency of this object is 1-5\%, 
which is smaller than the standard accretion efficiency. 
This indicates that Swift J0243.6+6124 is observed 
at a relatively large viewing angle 
within the range where the pulse emission is detected.
The magnetic field strength evaluated in the present study
is not inconsistent with previous studies
\citep{Tsygankov2018,Doroshenko2020}.
\citet{Tsygankov2018} estimated 
the upper limit for $B_{\rm NS}$ in Swift J0243.6+6124, 
$<6\times10^{12}~{\rm G}$,
from the fact that the transition to the propeller regime
was not detected even at a luminosity of $6\times10^{35}~{\rm erg~s^{-1}}$. 
\citet{Doroshenko2020} concluded that 
$B_{\rm NS}$ in Swift J0243.6+6124 is in $(3-9)\times10^{12}~{\rm G}$
and likely at the lower limit of this range.
We note that $B_{\rm NS}$ here is assumed not to change during the observation.
This assumption would be reasonable since 
it has been pointed out that 
the NS magnetic field strength 
does not change for a sufficiently long period, $>1~{\rm Myr}$,
based on the observations of some X-ray pulsars and the ULXP, NGC 1313 X-2
\citep{Makishima1999, Sathyaprakash2019}.
Although the NSs might have multipole magnetic fields,
we focus on the case of the dipole magnetic fields in this study.
We will discuss multipole fields later.

{\section{Discussions}\label{sec:discussion}}

\subsection{Geometrical beaming\label{sec:beaming}}
In our all models, the radiation is highly beamed 
by the outflows from the accretion disk. 
The amplification factor 
$\left<L_{\rm iso}\right>/\left<L_{\rm rad}\right>\equiv1/b$
exceeds 100 at maximum, 
which is higher than that indicated by \citet{King2019} (typically $1/b\sim1-10$) 
\citep[see also][]{Takahashi2017,Abarca2021}. 
The cause of this difference is probably the magnetic field strength. 
The difference will be smaller if we employ a strong $B_{\rm NS}$ 
by which $r_{\rm M}$ nearly equals $r_{\rm sph}$, 
although $r_{\rm M}$ is much smaller than $r_{\rm sph}$ in the present study. 
As $r_{\rm M}$ approaches $r_{\rm sph}$, 
the outflow rate would decrease since the outflows mainly occur 
in $r_{\rm M}< r< r_{\rm sph}$. 
In this case, the outflow can no longer effectively collimate the radiation, 
and therefore $1/b$ would decrease \citep{Abarca2021}. 
Although simulating $r_{\rm M}\sim r_{\rm sph}$ is difficult 
and beyond the scope of the present study (see also Section \ref{sec:future}), 
we plan to conduct such simulations in the future.

\subsection{Comparison with observations of 
NGC 5907 ULX1 and NGC 1313 X-2}

Here, 
we show, based on the present simulations,
that the ULXPs, 
NGC 5907 ULX1 and NGC 1313 X-2, 
are thought to be powered by the highly super-Eddington accretion
onto NSs with relatively weak magnetic fields.
In NGC 5907 ULX1, 
the mechanical power ($L_{\rm mec}^{\rm obs}$),
which is evaluated from the observed nebula emission of the ULX bubbles, 
is reported to $1.3\times10^{41}~{\rm erg~s^{-1}}$
\citep{Belfiore2020},
and is comparable to $\left<L_{\rm kin}\right>$ 
for the higher mass accretion rate model ({\tt\string B3e10D1}),
$\sim8\times10^{40}~{\rm erg~s^{-1}}$.
In addition, 
the isotropic luminosity $\left<L_{\rm ISO}\right>$ in 
model {\tt\string B3e10D1}
at $\theta\sim30^\circ$ or $150^\circ$
is almost the same as the observed X-ray luminosity,
$L_{\rm ISO}^{\rm obs}\sim2.2\times10^{41}~{\rm erg~s^{-1}}$
\citep{Israel2017a},
where the isotropic luminosity is 
calculated as 
$\left<L_{\rm ISO}\right>=-4\pi r_{\rm out}^2\left<R^r_t\right>$.
Thus,
NGC 5907 ULX1 probably has a mass accretion rate of 
$>10^3\dot{M}_{\rm Edd}$
and is viewed from at the polar angle 
20-30 degrees away from the $z$-axis.
The magnetic field strength of the NS is 
stronger than $10^{10}~{\rm G}$.
This is because 
$\left<L_{\rm kin}\right>$
and $\left<L_{\rm ISO}\right>$ of 
model {\tt\string B1e10D1\_np}
are almost the same as those of 
model {\tt\string B3e10D1},
but no accretion column appears in this model.

Also, $\left<L_{\rm kin}\right>$ 
for our slightly lower mass accretion models
({\tt\string B1e10D01}, {\tt\string B3e10D01}, 
and {\tt\string B3e10D02})
are comparable to $L_{\rm mec}^{\rm obs}$ in NGC 1313 X-2,
$\sim10^{40}~{\rm erg~s^{-1}}$
\citep{Pakull2008}.
In these models, the viewing angle,
at which $L_{\rm ISO}$ is consistent with 
$L_{\rm ISO}^{\rm obs}\sim(1.4-2.0)\times10^{40}~{\rm erg~s^{-1}}$,
is about 30 to 40 degrees measured from the z-axis.
Therefore, NGC 1313 X-2 would be almost a face-on object 
with a mass accretion rate of several $10^2\dot{M}_{\rm Edd}$.
Since no accretion column is formed 
in model {\tt\string B3e9D01\_np},
of which the kinetic and isotropic luminosity are 
similar to those of the above three models,
$B_{\rm NS}$ would be stronger than $3.3\times10^{9}~{\rm G}$.

According to our numerical models, 
both NGC 5907 ULX1 and NGC 1313 X-2 
would also have effectively optically thick outflows 
with $r_{\rm bb} > 100~{\rm km}$. 
Indeed, 
the thermal emission, probably originated from outflows, 
was reported in NGC 1313 X-2 \citep{Qiu2021}.
However, such emission is not detected in NGC 5907 ULX1.
This inconsistency can be resolved 
by conducting simulations of a large $r_{\rm M}$. 
If $B_{\rm NS}$ is so strong that $r_{\rm M}$ is comparable to $r_{\rm sph}$, 
then the outflow rate is thought to decrease, 
and thermal emission is not detected. 
In this case, $1/b$ decreases as discussed in Section \ref{sec:beaming}, 
and the observer's viewing angle should be smaller than that estimated above for 
$\left<L_{\rm ISO}\right>$ to be consistent with 
$L_{\rm ISO}^{\rm obs}\sim(1.4-2.0)\times10^{41}~{\rm erg~s^{-1}}$. 
However, increasing $B_{\rm NS}$ may change the intrinsic luminosity, 
and therefore simulations with a high $B_{\rm NS}$  is necessary to perform.
Post-processing radiative transfer simulations are needed
to obtain detailed radiation spectra and compare the observation
\citep{Kawashima2012,Kitaki2017,Narayan2017}.

On the other hand, 
no nebular has been detected around Swift J0243.6+6124.
The reason is that the object is a transient source. 
In order to form a nebula extending over $100~{\rm pc}$, 
an energy injection, $\sim10^{40}~{\rm erg~s^{-1}}$,
from the central object is needed to last for at least $10^{5}$ yr 
\citep{Pakull2008,Belfiore2020}.

\subsection{Ultraluminous supersoft sources}

Our simulations are consistent with 
the hypothesis proposed by \citet{Urquhart2016}
in which ULXs are observed 
as ultraluminous supersoft sources (ULSs) by an edge-on observer,
as far as the angular range of $\tau_{\rm eff}>1$ 
and its accretion rate dependence are concerned.
According to their model,
the optical depth of the outflows $\tau_{\rm eff}$ 
is larger than unity at $35^\circ<\theta<145^\circ$
for $\dot{M}_{\rm in}\sim100\dot{M}_{\rm Edd}$,
and 
the polar angle at which $\tau_{\rm eff} >1$ widens
as the mass accretion rate increases.
These features are also obtained in our simulations.
Indeed, simulated outflows are effectively optically thick
at polar angles greater than 40 degrees away 
from the $z$-axis for models 
{\tt\string B1e10D01} and {\tt\string B3e10D01}
(see Figures \ref{fig:figure6} and \ref{fig:figure7} for detail).
The mass accretion rates in these models are 
$500\dot{M}_{\rm Edd}$ and 
$720\dot{M}_{\rm Edd}$, respectively
(Table \ref{tab:table2}).
This angular range and the mass accretion rates 
do not so contradict the suggestion by \citet{Urquhart2016}
mentioned above.
Other models, in which the accretion rate
is $\sim 10^{2-3} \dot{M}_{\rm Edd}$
and the accretion columns form,
show approximately the same results.
In addition, the angular range where $\tau_{\rm eff} >1$ widens 
with increasing in the mass accretion rate
(see Figure \ref{fig:figure7}).
However, 
the blackbody temperature estimated from the simulation,
$\sim10^7~{\rm K}$, 
is higher than that observed in the ULSs, $\sim10^6~{\rm K}$.
Such a discrepancy might be resolved
by the simulations 
with the initial torus located far away
since the winds launched from the outer part 
of the accretion disks
work to expand the photosphere 
and lower the temperature there 
(we will discuss this point later).
If this is the case,
our simulations are consistent with the hypothesis 
in which ULSs are edge-on sources of ULXs.


\subsection{Why does the one-side accretion occur?}
\label{sec:one-side}

The accretion to only one of the poles 
(one-side accretion) occurs 
in models with a relatively large truncation radius
({\tt\string B3e9D001},  {\tt\string B1e10D001}, {\tt\string B1e10D002},
{\tt\string B3e10D001}, {\tt\string B3e10D01}, and
{\tt\string B3e10D02})
since the distorted magnetic field lines 
inhibit the gas accretion onto the opposite side.
Initially, the shape of the NS magnetic field 
is symmetric with respect to the equatorial plane.
Then, the magnetic field on the equatorial plane is perpendicular to the equatorial plane.
However, when the gas accumulates at the truncation radius and flows toward either pole, 
the magnetic field lines are distorted
(see magnetic field lines 
in the left panel of Figure \ref{fig:figure3}).
The magnetic field on the equatorial plane 
at the truncation radius
will be tilted with respect to the equatorial plane.
Since the gas tends to move to the side with the lower gravitational potential,
the gas flows in the direction 
in which the preceding gas accretes.
The accretion column on the opposite side 
is less likely to be formed.
This one-side accretion is more pronounced 
for the model with a large truncation radius 
({\tt\string B3e9D001}, {\tt\string B1e10D001}, 
{\tt\string B1e10D002}, {\tt\string B3e10D001}, 
{\tt\string B3e10D01}, and {\tt\string B3e10D02}).
In contrast, 
for the case where the truncation radius is very small,
the gas on the surface of the thick disk 
accretes to the opposite pole,
forming two accretion columns
(models 
{\tt\string B1e10D01}, {\tt\string B1e10D004}, 
{\tt\string B3e10D04}, and {\tt\string B3e10D1}).
The one-side accretion 
described above was also shown by some MHD simulations
\citep[see e.g.,][]{Lii2014,Takasao2022}.
%
%
We discuss the importance of three-dimensional simulations to study the detailed structure of accretion flow in the subsequent section.

\subsection{Future issues\label{sec:future}}

Three-dimensional simulations of non-axisymmetric accretion flows onto 
a rotating NS have yet to be performed.
In this study, we assume an axisymmetric structure,
in which the magnetic axis of the NS coincides with 
the rotation axis of the accretion disk,
and perform two-dimensional simulations.
However, the gas is thought to accrete onto the NS surface 
through multiple streams due to the magnetic Rayleigh–Taylor 
or interchange instability \citep{Kulkarni2008}. 
Furthermore, the pulse emission is produced 
by the misalignment between the magnetic axis and rotation axis of the NS.
Thus, we need to treat a non-axisymmetric structure
with three-dimensional simulations.
The misalignment of two axes 
might make the one-side accretion difficult to occur
since the gas in the accretion disk would preferentially 
accrete to the closest pole \citep[see e.g.,][]{Romanova2003}.
In addition, if the NS rotates as fast as millisecond pulsars, 
the outflow rate might increase 
due to the effective acceleration 
by a rapidly rotating magnetosphere of the NS \citep[e.g.,][]{Lovelace1999}.
Since the size of the photosphere increases as the mass outflow rate increases, 
the gas temperature decreases.
As a result, $r_{\rm bb}$ increases, 
and $B_{\rm NS}$ is larger than that estimated in Figure \ref{fig:figure8}. 
Simulations taking account of the rotation of the NS have been performed by 
\citet{Parfrey2017} and \citet{Das2022}. 


In our simulations, 
the gas-radiation interaction is switched off 
in the very vicinity of the rotation axis 
near the NS where $\sigma>\sigma_{\rm rad}$ 
(see Section \ref{sec:boundary}). 
This treatment would not affect the resulting size and temperature of the photosphere 
in Section \ref{sec:outflow}. 
In the present simulations, the radiatively driven outflows 
are launched from the accretion column base as well as the accretion disk 
\citep[see also][]{Abolmasov2022}. 
However, the opacity of free-free absorption of such outflows is quite small 
due to the low-density and high temperature. 
This leads to the smaller size of the photosphere for 
$\theta\lesssim\pi/6,5\pi/6\lesssim\theta$ compared with 
that formed by outflows from the accretion disk 
($r_{\rm th}\sim$ several $100~{\rm km}$, see red lines in Figure \ref{fig:figure6}). 
If we do not ignore the gas-radiation interaction near the rotation axis, 
the outflowing gas launched from the accretion column 
is more effectively accelerated by the radiation force near the rotation axis. 
This further reduces the gas density and size of the photosphere. 
Therefore, neglecting the gas-radiation interaction 
in the vicinity of the rotation axis near the NS 
does not affect the size and temperature of 
the photosphere presented in Section \ref{sec:outflow}.

We need to investigate the influence of the boundary condition. 
In this study, 
we assumed that the kinetic and thermal energy 
reaching the inner boundary (NS surface) 
is immediately converted into radiative energy. 
Under this boundary condition, 
the energy carried to the NS core is ignored. 
In reality, a fraction of the energy 
reaching the NS surface is converted into 
the thermal energy of the NS atmosphere. 
The heat flux from the atmosphere to the core occurs 
due to thermal conduction 
\citep[see][and references therein]{Wijnands2017}. 
Then, the radiative energy generated on the NS surface might decrease, 
leading to a decrease in radiative luminosity. 
Also, the accretion flow geometry might change 
if we set different boundary conditions (see Appendix \ref{sec:appendix1} for detail). 
Investigation of how the accretion flow depends on 
the boundary conditions is left as future work.

GR-RMHD simulations of large-scale accretion flows
are also needed to evaluate $r_{\rm bb}$ and $T_{\rm bb}$, more accurately.
It has been pointed out 
that the simulations in which the initial torus is located 
inside the spherization radius
overestimate the mass outflow rate
\citep[see][]{Kitaki2021,Yoshioka2022}.
Therefore, the accretion rate 
to reproduce effectively optically thick outflows 
would be higher than that estimated in the present study.
This point should be clarified in long-term simulations 
with the initial torus placed far enough away from the spherization radius.
In such simulations, the disk wind is expected to blow from the outer region of the accretion disks ($r \lesssim r_{\rm sph})$.
The blackbody radius would be larger, 
and the blackbody temperature would be smaller 
in the region near the equatorial plane. 
If this is the case, 
edge-on observers might identify the objects as the ULSs.

In the present study, 
we adopt a relatively weak magnetic field of the NS $3.3\times10^{9-10}$ G
compared to the typical X-ray pulsar $10^{11-13}$ G.
As $B_{\rm NS}$ increases while the accretion rate remains fixed, 
the amount of the outflowing gas probably reduces 
since the area where radiatively driven outflows mainly occur, 
$\pi (r_{\rm sph}^2-r_{\rm M}^2)$, decreases.
This leads the photosphere to be small 
and the gas temperature at the photosphere to be high.
It is expected that the blackbody radius 
tends to decrease with an increase in $B_{\rm NS}$.
In this case,
the magnetic field strength of Swift J0243.6+6124 
estimated in Section \ref{sec:outflow_Bfield} would decrease.
Although $r_{\rm bb}$ may depend on $B_{\rm NS}$ as described above, 
$r_{\rm bb}$ is a function of mass accretion rate only 
(see Table \ref{tab:table1} and equation \ref{eq:fitting})
in our results since simulations of $r_{\rm M}\ll r_{\rm sph}$ are performed.
In addition, it has been reported that when $B_{\rm NS}$ 
is strong enough that $r_{\rm M}>r_{\rm sph}$,
$r_{\rm M}$ is no longer dependent on $B_{\rm NS}$ \citep{Chashkina2017}.
We need to perform simulations for the strong $B_{\rm NS}$ case.

The M1 closure, which is employed in the present study, 
leads to unphysical solutions 
when the system is optically thin and the radiation fields are anisotropic
\citep[see e.g.,][]{Ohsuga2016}.
In our simulations, 
especially in the models where the one-side accretion occurs, 
the radiation fields are quite anisotropic.
In order to accurately calculate the radiation fields,
we have to solve the radiative transfer equation.
Such radiation MHD simulations,
in which the radiative transfer equation is solved, 
are performed by 
\citet[][]{Jiang2014},
\citet[][]{Ohsuga2016},
\citet[][]{Asahina2020},
\citet[][]{Zhang2022},
and \citet[][]{Asahina2022}.

The modeling of Swift J0243.6+6124 
considering multipolar magnetic field components 
is to be further investigated.
Although we constrain the dipole magnetic field strength 
at the NS surface,
$3\times10^{11}~{\rm G}<B_{\rm NS}<4\times10^{12}~{\rm G}$,
in the present work
\citep[see also][]{Tsygankov2018,Doroshenko2020}, 
a cyclotron resonance scattering feature (CRSF) 
corresponding to $1.6\times10^{13}~{\rm G}$ 
was reported by \citet{Kong2022}.
This discrepancy is resolved 
if CRSF originates from multipole magnetic fields,
as already suggested by some authors \citep[see e.g.,][]{Israel2017a}.
In the case that 
the multipole magnetic field component 
is dominant over the dipole component,
the accretion flow geometry is expected to 
be more complex \citep[][]{Long2007,Das2022}, 
and the position of the magnetospheric radius changes.
We plan to perform the GR-RMHD simulations 
of super-Eddington accretion flows onto the
NS with multipole magnetic fields.

Our models might exhibit a high polarization degree, 
as recently observed in the X-ray binary 
\citep{Veledina2023,Ratheesh2023}. 
The hard X-ray photons produced in the accretion column base 
are scattered by the inner wall of funnel, 
which consists of the effectively optically thick outflows, 
and pass through the low-density region near the rotation axis. 
If such photons are observed, a high polarization degree might be detected. 
Indeed, \citet{Ratheesh2023} reported 
that the polarization degree increases from approximately 6\% at 2 keV to 10\% at 8 keV.
Post-process polarized radiative transfer simulations are needed 
to compare our models with such observations.

\section{Conclusions}

We performed two-dimensional axisymmetric GR-RMHD simulations of super-Eddington flows around NSs with a dipole magnetic field 
for modeling the galactic ULXP, Swift J0243.6+6124.
In our simulations, 
the accretion columns near the magnetic poles, 
the super-Eddington disk outside the magnetospheric radius, 
and outflows launched from the disk appear.
If the magnetospheric radius (truncation radius)
is smaller than the spherization radius,
the outflows are generated 
since the radiation force mainly drives the outflows
inside the spherization radius.
When the accretion rate is large enough
while satisfying the above condition,
effectively optically thick outflows are launched from the disk
and would be responsible for the blackbody radiation.
The blackbody temperature of the 
effectively thick outflow, 
$\sim 10^7$ K, is roughly consistent with the observations by \citet{Tao2019}.
The blackbody radius increases with an
increase in the mass accretion rate 
and agrees with observations, $100-500$ km,
when the accretion rate is about $(130-1200)\dot{M}_{\rm Edd}$.
Since the blackbody radiation was detected in two 
observations with $\dot{P}$ of
$-2.22\times10^{-8}~{\rm s~s^{-1}}$ and 
$-1.75\times10^{-8}~{\rm s~s^{-1}}$, 
but not in another with
$\dot{P}\sim-6.8 \times10^{-9}~{\rm s~s^{-1}}$,
the surface magnetic field strength of the NS in
Swift J0243.6+6124 is limited to be
$3\times10^{11}~{\rm G} \lesssim 
B_{\rm NS} \lesssim 4\times10^{12}~{\rm G}$.
The accretion rate is evaluated to be 
$(200-500)\dot{M}_{\rm Edd}$
when the blackbody radiation is detected and 
$(60-100)\dot{M}_{\rm Edd}$ 
when the blackbody radiation is not observed.
Our results support the hypothesis 
that the super-Eddington phase 
in the 2017-2018 giant outburst of Swift J0243.6+6124
is powered by super-Eddington accretion flows onto a magnetized NS.

\begin{acknowledgments}
We would like to thank an anonymous reviewer for fruitful comments.
This work was supported by JSPS KAKENHI Grant Numbers 
JP21J21040 (A.I.), 
JP21H04488, JP18K03710 (K.O.), 
JP20K11851, JP20H01941, JP20H00156 (H.R.T) 
and JP18K13591 (Y.A.). 
A part of this research has been funded by the MEXT as 
”Program for Promoting Researches on the Supercomputer Fugaku” 
(Toward an unified view of the universe: 
from large scale structures to planets, JPMXP1020200109) 
(K.O., H.R.T., Y.A., and A.I.), 
and by Joint Institute for Computational Fundamental Science (JICFuS, K.O.).
Numerical computations were performed with computational resources 
provided by the Multidisciplinary Cooperative Research Program 
in the Center for Computational Sciences, University of Tsukuba, 
Oakforest-PACS operated by the Joint Center for 
Advanced High-Performance Computing (JCAHPC), 
Cray XC 50 at the Center for Computational Astrophysics (CfCA) of 
the National Astronomical Observatory of Japan (NAOJ), 
the FUJITSU Supercomputer PRIMEHPC FX1000 
and FUJITSU Server PRIMERGY GX2570 (Wisteria/BDEC-01) 
at the Information Technology Center, The University of Tokyo. 
\end{acknowledgments}

%






\appendix

\begin{figure*}[htb]
\plotone{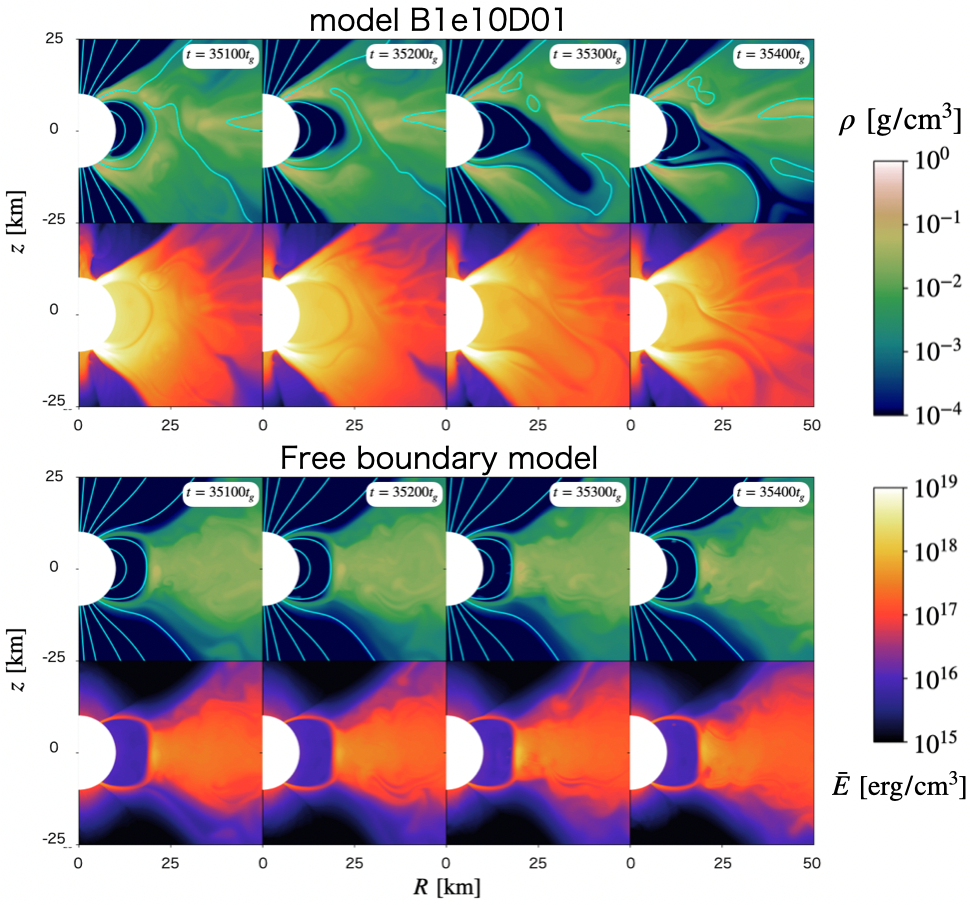}
\caption{
Time evolution of 
the mass density and  
the radiation energy density.
The colors show the gas density and radiation energy density.
The cyan lines represent the magnetic field lines.
The top panel is the results of model {\tt\string B1e10D01}.
The results of the model, 
where the free boundary condition is adopted for radiative flux
(free boundary model),
are presented in the bottom panel.
All the same initial parameters as {\tt\string B1e10D01} 
are adopted in the free boundary model
except for the boundary condition of the radiative flux.
\label{fig:figure9}}
\end{figure*}

{\section{Dependence of the accretion flows on the boundary conditions} \label{sec:appendix1}}
Here, we show the dependence of the accretion flows
on the inner boundary condition of the radiative flux.
Figure \ref{fig:figure9} shows 
the time evolution of the gas density 
and radiation energy density
with different boundary conditions.
The top panel is the result of model {\tt\string Be10D01}
where the radiative flux at the NS surface is set to zero.
On the other hand, 
the results of the model 
in which the radiative flux is set according to 
the free boundary condition are presented in the bottom panel
(free boundary model).
Cyan lines show the magnetic field lines.
We can see that the low-density void region is formed 
between two accretion columns 
($10~{\rm km}<r<13~{\rm km}$ and $50^\circ<\theta<130^\circ$). 
A part of the photons emitted from the accretion columns 
is transported to the void region. 
These photons push the gas outward and 
the void is suffered from expansion 
(see, the right top panel of Figure \ref{fig:figure9}). 
This void repeatedly contracts and expands. 
The period of repeated expansion is about 
$\sim 500t_g \simeq 10^{-3}~{\rm s}$. 
However, the variation in the mass accretion rate 
at the NS surface due to the expansion is negligibly small.
Such a structure can not be seen in the free boundary model.
This is because the photon can freely penetrate the NS surface,
and the outward radiation force considerably becomes small
compared with model {\tt\string B1e10D01}.

We also find that 
the thickness of the accretion column 
in the free boundary model
is thinner than that of model {\tt\string B1e10D01}.
It may originate from the decrease in the radiation energy
of the accretion column due to the free boundary condition.
It can be seen that 
the radiation energy at the base of the accretion column 
in model {\tt\string B1e10D01}
is about one hundred times larger than 
that in the free boundary model.

{\section{Analytic formula of magnetospheric radius and spin-up rate} \label{sec:appendix2}}

\citet{Takahashi2017} derived
the analytical formulas of 
the magnetospheric radius (\ref{eq:rM_analize}), 
and spin-up rate (\ref{eq:Pdot_analize}),
applying the self-similar solutions 
for the slim disk proposed by \citet{Watarai1999}.
Here we summarize the derivations of these formulas.

In the slim disk model, 
the energy balances are 
$Q_{\rm adv}^-=Q^+_{\rm vis}+Q^-_{\rm rad}=fQ_{\rm vis}^+$ 
\citep{Narayan1994},
where $Q_{\rm adv}^-$ is the advected energy,
$Q_{\rm vis}^+$ is the viscous-dissipated energy,
$Q_{\rm rad}^-$ is the radiative cooling energy,
and $f$ represents the fraction of the advective cooling 
to the viscous-dissipated energy.
Using the height-integrated radiation pressure 
described in equation 12 in \citet{Watarai1999} and
the scale height of the accretion disk $H$, presented in 
equation 11 in \citet{Watarai1999},
we get
\begin{eqnarray}
p_{\rm rad}
=\Pi_{\rm rad}/2H
=\frac{\dot{M}}{4\pi\alpha}\frac{c_3^{1/2}}{c_1}\sqrt{\frac{GM_{\rm NS}}{r^5}},
\end{eqnarray}
where $\alpha$ is the viscous parameter \citep{Shakura1973}, 
$c_1$ and $c_3$ is the constant depenting on 
$\Gamma$, $f$ and $\alpha$ \citep[see][for detail]{Watarai1999}.
In this study, we take $\Gamma=1.5$, $f=0.1$ for the analytical formula.
The reason why we do not employ $\Gamma=5/3$ is that
the rotational velocity of the disk is zero with $\Gamma=5/3$ \citep{Narayan1994}.
The magnetic pressure at radius $r$ by the dipole magnetic field is
$p_{\rm mag}=B_{\rm NS}^2/8\pi(r_{\rm NS}/r)^6$.
The magnetospheric radius $r_{\rm M}$ is defined as 
the radius where $p_{\rm mag}$ balances with $p_{\rm rad}$:
\begin{eqnarray}
r_{\rm M}
&=&
\left(
\frac{1}{8\pi}\frac{c_1}{c_3^{1/2}}
\frac{c\kappa_{\rm sca}}{G^{3/2}}
\right)^{2/7}\nonumber\\
&\times&
\alpha^{2/7}
\left(
\frac{\dot{M}_{\rm in}}{\dot{M}_{\rm Edd}}
\right)^{-2/7}
B_{\rm NS}^{4/7}
M_{\rm NS}^{-3/7}
r_{\rm NS}^{12/7}.
\label{eq:rM_appendix}
\end{eqnarray}

Next we give the derivation of the spin-up rate.
The spin-up rate can be calculated by assuming 
that the Keplerian angular momentum at $r_{\rm M}$
is transported to the NS without dissipation.
In this case, spin-up rate is represented by
$\dot{P}=-\dot{M}l({r_{\rm M}})/M_{\rm NS}l_{\rm NS}$
\citep{Shapiro1986},
where $l=\sqrt{GM_{\rm NS}r}$
is the specific Keplerian angular momentum.
Using (\ref{eq:rM_appendix}), 
the spin-up can be calculated as follows:
\begin{eqnarray}
\dot{P}
&=&
-\frac{2G^{3/2}}{c\kappa_{\rm sca}}
\left(
\frac{1}{8\pi}\frac{c_1}{c_3^{1/2}}
\frac{c\kappa_{\rm sca}}{G^{3/2}}
\right)^{1/7}\nonumber\\
&\times&
\alpha^{1/7}
\left(\frac{\dot{M}_{\rm in}}{\dot{M}_{\rm Edd}}\right)^{6/7}\nonumber\\
&\times&
B_{\rm NS}^{2/7}
M_{\rm NS}^{2/7}
r_{\rm NS}^{-8/7}
P^2.
\end{eqnarray}

{\section{Equations of motion in general relativity} \label{sec:appendix3}}

Here, we derive the equations of motion, 
which are used in this study, according to \citet{Mihalas1984}.
Equation (\ref{eq:momentum_cons_gas}) can be transformed to
\begin{eqnarray}
\partial_\mu T^{\mu\nu}=
-T^{\mu\nu}\frac{\partial_\mu\sqrt{-g}}{\sqrt{-g}}
-\Gamma^\nu_{\alpha_\mu}T^{\alpha\mu}
+G^\nu.
\label{eq:momentum_cons_appendix}
\end{eqnarray}
Subtracting the time component of (\ref{eq:momentum_cons_appendix}) 
multiplied by $u^i/u^t$ from the spatial component 
of (\ref{eq:momentum_cons_appendix}), 
the following relation can be obtained:
\begin{eqnarray}
\partial_\mu T^{\mu i}&-&\frac{u^i}{u^t}\partial_\mu T^{\mu t}\nonumber\\
=
&-&\left(T^{\mu i}-\frac{u^i}{u^t}T^{\mu t}\right)
\frac{\partial_\mu\sqrt{-g}}{\sqrt{-g}}\nonumber\\
&-&
\left(\Gamma^i_{\alpha\mu}-\frac{u^i}{u^t}\Gamma^t_{\alpha\mu}\right)
T^{\alpha\mu}+G^i-\frac{u^i}{u^t}G^t.
\label{eq:calc_appendix1}
\end{eqnarray}
We define the relativistic enthalpy, $w=\rho+p_{\rm gas}+e+b^2$, 
and equation (\ref{eq:calc_appendix1}) can be written as
\begin{eqnarray}
\partial_\mu T^{\mu i}-\frac{u^i}{u^t}\partial_\mu T^{\mu t}
&=&wu^\mu\left[
\partial_\mu\left\{e^i_{(\alpha)}\right\}
-\frac{u^i}{u^t}\partial_\mu\left\{e^t_{(\alpha)}\right\}
\right]u^{(\alpha)}\nonumber\\
&+&w\left[
e^i_{(\alpha)}-\frac{u^i}{u^t}e^t_{(\alpha)}
\right]\frac{du^{(\alpha)}}{d\tau}\nonumber\\
&+&\partial_\mu\left\{(p_{\rm gas}+p_{\rm mag})g^{\mu i}\right\}\nonumber\\
&-&\frac{u^i}{u^t}\partial_\mu
\left\{(p_{\rm gas}+p_{\rm mag})g^{\mu t}\right\}\nonumber\\
&-&\partial_\mu\left(b^\mu b^i\right)
+\frac{u^i}{u^t}\partial_\mu\left(b^\mu b^t\right),
\label{eq:bardeen}
\end{eqnarray}
where 
${e^{(\mu)}}_{\alpha}$ is the orthonormal tetrad 
that transforms vectors in the laboratory frame 
(e.g., Boyer–Lindquist coordinate) 
to the zero angular momentum observer frame 
\citep[ZAMO frame,][]{Bardeen1972},
$u^{(\mu)}={{e^{(\mu)}}_{\alpha}}u^{\alpha}$ 
is the four-velocity of the gas in the ZAMO frame,
${e^{\mu}}_{(\alpha)}$ is the inverse matrix of ${e^{(\mu)}}_{\alpha}$,
and $d/d\tau=u^\mu\partial_\mu$ is the derivative of proper time, $\tau$
\citep{Mihalas1984}.
Now using equation (\ref{eq:calc_appendix1}) and (\ref{eq:bardeen}),
we get
\begin{eqnarray}
&w&\left[
e^i_{(\alpha)}-\frac{u^i}{u^t}e^t_{(\alpha)}
\right]
\frac{du^{(\alpha)}}{d\tau}\nonumber\\
&=&
-wu^\mu\left[
\partial_\mu\left\{e^i_{(\alpha)}\right\}
-\frac{u^i}{u^t}\partial_\mu\left\{e^t_{(\alpha)}\right\}
\right]u^{(\alpha)}\nonumber\\
&-&\partial_\mu\left\{(p_{\rm gas}+p_{\rm mag})g^{\mu i}\right\}
+\frac{u^i}{u^t}\partial_\mu
\left\{(p_{\rm gas}+p_{\rm mag})g^{\mu t}\right\}\nonumber\\
&+&\partial_\mu\left(b^ib^\mu\right)
-\frac{u^i}{u^t}\partial_\mu\left(b^tb^\mu\right)\nonumber\\
&-&\left(T^{\mu i}-\frac{u^i}{u^t}T^{\mu t}\right)
\frac{\partial_\mu\sqrt{-g}}{\sqrt{-g}}\nonumber\\
&-&\left(\Gamma^i_{\alpha\mu}-\frac{u^i}{u^t}\Gamma^t_{\alpha\mu}\right)T^{\alpha\mu}
+G^i-\frac{u^i}{u^t}G^t.
\end{eqnarray}

We calculate the equation of motion in the $r$-direction.
In order to obtain a concrete expression, 
it is necessary to fix the space-time.
We consider the Boyer-Lindquist metric with a spin parameter of zero.
In this case, 
all nondiagonal components of ${{e^{(\mu)}}_{\alpha}}$ are zero.
Assuming steady ($\partial_t=0$) and axisymmetric ($\partial_\phi=0$) flow,
the following equation of motion in the radial direction is obtained:
\begin{eqnarray}
f_{\rm grav}^{(r)}
+f_{\rm thermal}^{(r)}
+f_{\rm \theta,cent}^{(r)}
+f_{\rm \phi,cent}^{(r)}
+f_{\rm rad}^{(r)}
+f_{\rm mag}^{(r)}
+f_{\rm adv}^{(r)}
=0
\label{eq:eom_appendix1}
\end{eqnarray}
where 
$f_{\rm grav}^{(r)}$, $f_{\rm thermal}^{(r)}$, $f_{\rm \theta,cent}^{(r)}$, 
$f_{\rm \phi,cent}^{(r)}$, $f_{\rm rad}^{(r)}$, $f_{\rm mag}^{(r)}$, and $f_{\rm adv}^{(r)}$
describe 
the gravitational force, 
thermal force, 
centrifugal force due to the poloidal motion,
centrifugal force due to the toroidal motion,
radiation force,
Lorentz force, 
and advection force, respectively.
These force can be described as follows:
\begin{eqnarray}
f_{\rm grav}^{(r)}
&=&\frac{M_{\rm NS}}{r^2}w\left(u^tu_t+u^ru_r\right),\\
f_{\rm thermal}^{(r)}
&=&-\left(1-\frac{2M_{\rm NS}}{r}\right)\partial_r p_{\rm gas},\\
f_{\rm \theta,cent}^{(r)}
&=&\left(1-\frac{2M_{\rm NS}}{r}\right)w\frac{u^\theta u_\theta}{r},\\
f_{\rm \phi,cent}^{(r)}
&=&\left(1-\frac{2M_{\rm NS}}{r}\right)w\frac{u^\phi u_\phi}{r},\\
f_{\rm rad}^{(r)}&=&G^r-\frac{u^r}{u^t}G^t,\\
f_{\rm mag}^{(r)}
&=&
-\left(1-\frac{2M_{\rm NS}}{r}\right)\partial_r
\left(\frac{b^2}{2}\right)\nonumber\\
&+&\partial_\mu\left(b^\mu b^r\right)
-\frac{u^r}{u^t}\partial_\mu\left(b^\mu b^t\right)\nonumber\\
&+&\left(\frac{2b^r}{r}+\frac{b^\theta}{\tan\theta}\right)
\left(b^r-\frac{u^r}{u^t}b^t\right)\nonumber\\
&-&\frac{M_{\rm NS}}{r}\frac{b^t b_t+b^r b_r}{r}\nonumber\\
&-&\left(1-\frac{2M_{\rm NS}}{r}\right)
\frac{b^\theta b_\theta+b^\phi b_\phi}{r}\nonumber\\
&-&\frac{2M_{\rm NS}}{r^2}\frac{u^r}{u^t}b^t b_r,\\
f_{\rm adv}^{(r)}
&=&-w\gamma
\left(1-\frac{2M_{\rm NS}}{r}\right)^{1/2}
\left(u^r\partial_r+u^\theta\partial_\theta\right)v^{(r)}
\end{eqnarray}
where $\gamma=u^{(t)}$ is the Lorentz factor 
in the ZAMO frame (i.e. statistic observer frame),
and $v^{(r)}=u^{(r)}/r^{(t)}$.
The equation of motion in the $\theta$-direction 
can also be obtained by the same procedure.
\begin{eqnarray}
f_{\rm inertial}^{(\theta)}
+f_{\rm thermal}^{(\theta)}
+f_{\rm \phi,cent}^{(\theta)}
+f_{\rm rad}^{(\theta)}
+f_{\rm mag}^{(\theta)}
+f_{\rm adv}^{(\theta)}=0,
\label{eq:eom_appendix2}
\end{eqnarray}
where
$f_{\rm inertial}^{(\theta)}$ is the inertial force caused 
by the motion in the direction of $\theta$.
These force are written as follows:
\begin{eqnarray}
f_{\rm inertial}^{(\theta)}
&=&-wu_ru^\theta
\left(1-\frac{3M_{\rm NS}}{r}\right),\\
f_{\rm thermal}^{(\theta)}
&=&-\frac{1}{r}\partial_\theta p_{\rm gas},\\
f_{\rm \phi,cent}^{(\theta)}
&=&w\frac{u^\phi u_\phi}{r\tan\theta},\\
f_{\rm rad}^{(\theta)}&=&r\left(G^\theta-\frac{u^\theta}{u^t}G^t\right),\\
f_{\rm mag}^{(\theta)}
&=&-\frac{1}{r}\partial_\theta \left(\frac{b^2}{2}\right)
+r\partial_\mu\left(b^\mu b^\theta\right)\nonumber\\
&-&r\frac{u^\theta}{u^t}\partial_\mu\left(b^\mu b^t\right)
+r\left(\frac{2b^r}{r}+\frac{b^\theta}{\tan\theta}\right)
\left(b^\theta-\frac{u^\theta}{u^t}b^t\right)\nonumber\\
&+&2b^rb^\theta-\frac{b^\phi b_\phi}{r\tan\theta}
-\frac{u^\theta}{u^t}\frac{2M}{r}b^tb_r,\\
f_{\rm adv}^{(\theta)}
&=&-w\gamma
\left(u^r\partial_r+u^\theta\partial_\theta\right)v^{(\theta)},
\end{eqnarray}
where $v^{(\theta)}=u^{(\theta)}/u^{(t)}$.
The relativistic equations of motion derived by this method
reduce to the Newtonian equations of motion for
$\mathcal{O}((w-\rho)/\rho)\rightarrow0$, 
$\mathcal{O}(v/c)\rightarrow0$,
and $\mathcal{O}(M/r)\rightarrow0$.

\bibliography{library,mybook}
\bibliographystyle{aasjournal}

\end{CJK*}
\end{document}